\documentclass[fleqn,usenatbib]{mnras}

\usepackage{newtxtext,newtxmath}

\usepackage[T1]{fontenc}

\DeclareRobustCommand{\VAN}[3]{#2}
\let\VANthebibliography\thebibliography
\def\thebibliography{\DeclareRobustCommand{\VAN}[3]{##3}\VANthebibliography}

\usepackage{graphicx}
\usepackage{amsmath}

\usepackage{amssymb}
\usepackage{booktabs}
\usepackage{multirow}
\usepackage{multicol}
\usepackage{xcolor}

\title[Deep Learning for Asteroid Streak Detection]{Discovering Faint and High Apparent Motion Rate Near-Earth Asteroids Using A Deep Learning Program}

\author[F. Wang et al.]{
Franklin Wang$^{1,2}$, \thanks{E-mail: wfrankw9@gmail.com}
Jian Ge$^{3}$, \thanks{E-mail: jge@shao.ac.cn}
Kevin Willis$^{1}$
\\
$^{1}$Science Talent Training Center, Gainesville, FL 32606, USA\\
$^{2}$Palo Alto Senior High School, 50 Embarcadero Road, CA 94301, USA\\
$^{3}$Division of Optical Astronomical Instrumentation, Shanghai Astronomical Observatory, 80 Nandan Road, 200030, China
}

\date{Accepted 2022 August 16. Received 2022 August 05; in original form 2022 June 03}

\pubyear{2022}

\begin{document}
\label{firstpage}
\pagerange{\pageref{firstpage}--\pageref{lastpage}}
\maketitle

\begin{abstract}
Although many near-Earth objects have been found by ground-based telescopes, some fast-moving ones, especially those near detection limits, have been missed by observatories. We developed a convolutional neural network for detecting faint fast-moving near-Earth objects. It was trained with artificial streaks generated from simulations and was able to find these asteroid streaks with an accuracy of 98.7\% and a false positive rate of 0.02\% on simulated data. This program was used to search image data from the Zwicky Transient Facility (ZTF) in four nights in 2019, and it identified six previously undiscovered asteroids. The visual magnitudes of our detections range from $\sim$19.0-20.3 and motion rates range from $\sim$6.8-24 deg/day, which is very faint compared to other ZTF detections moving at similar motion rates. Our asteroids are also $\sim$1 - 51 m diameter in size and $\sim$5 - 60 lunar distances away at close approach, assuming their albedo values follow the albedo distribution function of known asteroids. The use of a purely simulated dataset to train our model enables the program to gain sensitivity in detecting faint and fast-moving objects while still being able to recover nearly all discoveries made by previously designed neural networks which used real detections to train neural networks. Our approach can be adopted by any observatory for detecting fast-moving asteroid streaks.
\end{abstract}


\section{Introduction}

Near-Earth objects (NEOs), which consist of asteroids and comets that have a perihelion distance of less than 1.3 AU, are of significant interest to the general public due to the impacts with Earth. Most notably, around 65 million years ago, an  impact from a 5 to 15 km diameter asteroid is believed to have led to the extinction of the dinosaurs \citep{Alvarez_etal_1980}. There have however, also been notable asteroid collisions within the last century. For instance, the $\sim$19 m Chelyabinsk meteor in 2013 exploded over a city in Russia, injuring many people, and the $\sim$40 m Tunguska meteor in 1908 exploded above a Siberian forest, flattening 80 million trees and generating an explosion at least ten times larger than the Chelyabinsk meteor. Based on past impacts, asteroids 19 m and larger are predicted to collide with Earth roughly every 25 years, and thus pose a threat to Earth \citep{brown_500-kiloton_2013}. 

Because of the danger NEOs pose, NASA’s goal, set in 2005, was to be able to find 90\% of asteroids above 140 m in diameter before 2020 \citep{nasa2007}. However, so far only around 30\% of the predicted 25,000 near Earth asteroids between 140 m and 1 km in diameter have been found. This is extremely concerning, as asteroids below 140 m have even less detection completeness and still pose a significant risk to the Earth. For instance, just 0.1\% of the predicted number of near Earth asteroids between 19 m and 44 m have been discovered \citep{b612_foundation_asteroid_2020}.

Being able to discover NEOs also allows us to better understand solar system formation, such as through sample collection missions like OSIRIS-REx \citep{lauretta_osiris-rex_2017} and through improved understandings of asteroid parameter distributions \citep{demeo_solar_2014}.

Asteroids with a high enough apparent motion will leave ``streaks" on telescope exposures as they move significantly during a telescope exposure. By specifically searching for these streaks, we can improve our ability to detect fast moving asteroids as streaks allow us to measure the approximate direction and speed of travel, making it much easier to link detections together. 

There have been many examples of streak detection algorithms that leverage conventional computer vision techniques. For instance, \citet{Nir_2018} uses an approach based on the Radon transform to find asteroid streaks and \citet{dawson_blind_2016} uses an approach that leverages signal-matched-filters. These works are able to effectively detect very dim streaks, but they are intended for use on streaks hundreds of pixels in length, and do not work as well for shorter streaks \citep{Nir_2018,dawson_blind_2016}. 

To overcome this issue, one strategy that has been recently adopted is the use of Convolutional Neural Networks (CNNs), a type of deep learning model that is highly effective for image classification. CNNs have been used by the Asteroid Terrestrial-impact Last Alert System (ATLAS) in \citet{Chyba_Rabeendran_2021} and the Zwicky Transient Facility (ZTF) in \citet{deepstreaksduev} to reduce the number of possible asteroid streak candidates that human scanners have to review. 

However, both of these methods have the drawback of incorporating real asteroid streak images in their datasets, which limited the size of their training datasets and likely biased them towards bright streaks. This is because asteroid streaks can be somewhat rare and it is much easier to find bright asteroid streaks than fainter ones \citep{Ye_2019}. For instance, the dataset used to train the CNN classifier for ATLAS only contains around 500 streaks and the dataset for ZTF only has roughly 15,000 streaks, even though it combines both simulated streaks and real data \citep{Chyba_Rabeendran_2021,deepstreaksduev}. The use of real streaks as opposed to simulated streaks also makes it difficult for these approaches to be applied to other observatories, who may not already have access to a large labelled dataset of streaks.

To overcome these limitations, this work strives to use a large, purely simulated dataset of asteroid streaks that focuses on fainter, harder to detect asteroid streaks. While using a purely simulated dataset for asteroid streak detection has been pursued by \citet{Lieu_etal_2019} in preparation for the future Euclid mission, their model has difficulty detecting short streaks (only 50\% accuracy for streaks 15 to 30 pixels in length) and has not yet been applied to real datasets. Because ZTF has released full nights of data to the public from past years, and to provide a comparison between this work and the research conducted by ZTF, we utilize publicly released data from ZTF spanning from 2018 to 2019.

In this paper, we describe our data collection and simulations, neural network, false positive reduction, and full detection pipeline methodology in section \ref{sec:method}; our results and discoveries in section \ref{sec:results}; and discussion and summary in section \ref{sec:discussion}. 

\section{Methodology} \label{sec:method}

The most optimal way to detect asteroid streaks would be to simulate every possible streak -- all the possible orientations, lengths, and brightnesses -- and see how closely every region in an image matches that streak \citep{Nir_2018}. This is, obviously, not feasible due to the high computational power required. Instead, our approach is to generate a large dataset of simulated streaks and then create a classifier using a convolutional neural network to distinguish between images with or without a true streak. 

\subsection{Dataset Collection and Simulations}

\subsubsection{Real Streak Data Collection}
\begin{figure}
    \centering
    \includegraphics[width=0.4\textwidth]{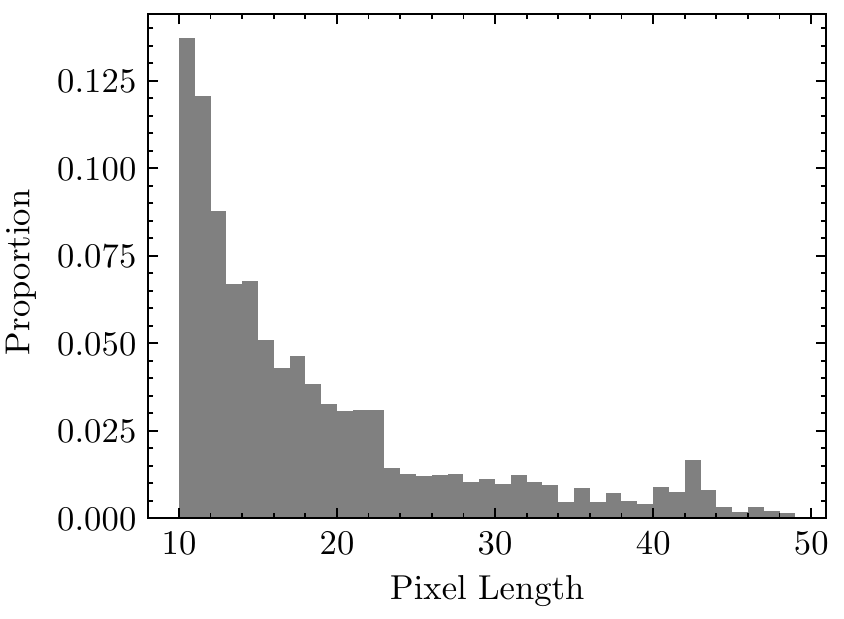}
    \caption{Normalized histogram of lengths of asteroid streaks from our set of real asteroid streaks.}
    \label{fig:lenshist}
\end{figure}

\begin{figure}
    \centering
    \includegraphics[width=0.4\textwidth]{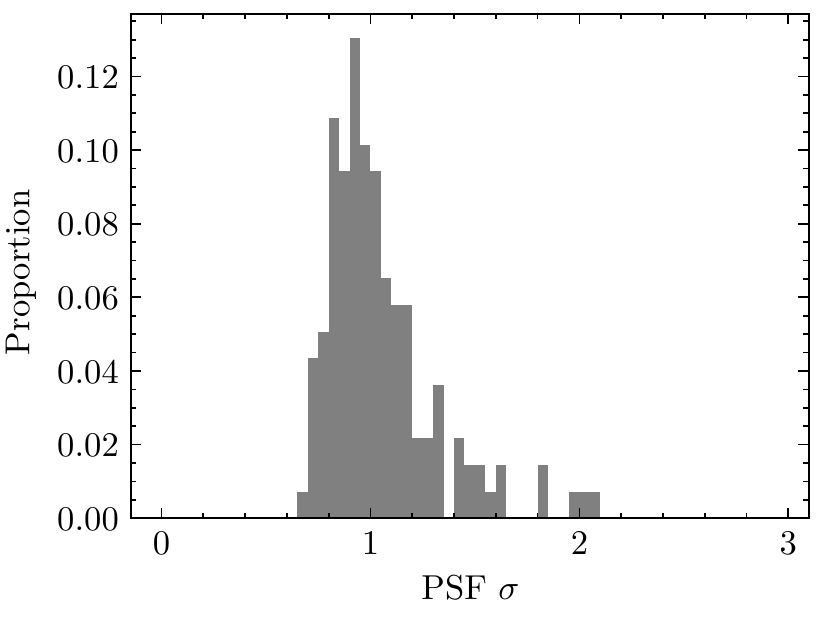}
    \caption{Normalized histogram of the Gaussian PSF $\sigma$ parameter (see Equation \ref{eq:flux}) of asteroid streaks from our set of real asteroid streaks.}
    \label{fig:widthshist}
\end{figure}

To ensure our dataset of simulated streaks adequately matches real asteroid streaks, we collect a small dataset of real streaks by first collecting a list of the asteroids which made a close approach to Earth of within 30 lunar distances during the ZTF survey's timeframe using the NASA CNEOS NEO Earth Close Approaches database\footnote{\url{https://cneos.jpl.nasa.gov/ca/}}. We then use the Moving Object Search Tool provided by IRSA to extract all occurrences of these asteroids within ZTF's survey images where the asteroid will be moving fast enough to leave a trail longer than 10 pixels in length \citep{YauMOST}. From this we have found the distributions for the streak lengths and widths described by the Gaussian PSF $\sigma$ parameter (see Equation \ref{eq:flux}) shown in Figures \ref{fig:lenshist} and \ref{fig:widthshist}. These distributions are used in the next section to help sample streak parameters, although we later decided to include shorter streaks less than 10 pixels in length in our dataset to ensure that shorter streaks could also be detected. This dataset of real streaks is not used to directly train the model. Instead, the training dataset we use is completely simulated, which helps ensure there is no bias due to real streaks tending to be brighter.

\subsubsection{Streak Simulation}
In order to generate simulated images of asteroid streaks, we model an image from an asteroid as being generated according to a 2D Gaussian PSF, which has been shown to accurately model asteroid streaks in \citet{veres_improved_2012}. Because the asteroid moves during the 30-second exposure, we can calculate the pixel intensities by convolving a line of pixels by a 2D Gaussian kernel, which results in the following equation from \citet{Ye_2019} for the flux at position $x, y$ for a streak with angle to the horizontal $\theta$, PSF width $\sigma$, length $L$, center $x_0$, $y_0$ and total flux $\Phi$
\begin{equation}\label{eq:flux}
    F(x, y) = \frac{\Phi}{2\sigma L\sqrt{2\pi}}\exp\left(-\frac{y'^2}{2\sigma^2}\right)\left(\text{erf}\left(\frac{x'+\frac{L}{2}}{\sigma\sqrt{2}}\right)-\text{erf}\left(\frac{x'-\frac{L}{2}}{\sigma\sqrt{2}}\right)\right)
\end{equation}
where we define $\text{erf}(z)$ as 
\begin{equation}
    \text{erf}(z)=\frac{2}{\pi}\int^z_0 e^{t^2}dt
\end{equation}
and $x'$ and $y'$ as
\begin{align}
    x'&=(x-x_0)\cos\theta + (y-y_0)\sin\theta\\
    y'&=-(x-x_0)\sin\theta + (y-y_0)\cos\theta
\end{align}

ZTF's data consists of science images, which are individual observed data frames; reference images, which are combined images containing static background stars; and difference images, which consist of transients generated by subtracting the science images from the reference images. In order to generate a training sample, a streak is generated using the 2D Gaussian PSF model and then implanted into a randomly selected 80 by 80 pixel region of a science image. The corresponding 80 by 80 pixel region of the reference image is also extracted. This gives us a science-reference pair of images, which will be input into a convolutional neural network. Though past work by ZTF \citep{deepstreaksduev} and ATLAS \citep{Chyba_Rabeendran_2021} only input difference image cutouts into their machine learning models, we have found that by using science and reference images, more information is conserved as image subtraction causes information from the science and reference images to be lost. Science and reference image cutouts are also used by more recent works by ZTF such as \citet{duevrb} and \citet{Duev_2021}. 

To generate a full dataset of asteroid streaks, we repeat the streak creation and implanting process multiple times by randomly sampling the required parameters. Firstly, the rotations are sampled randomly from 0 to 360 degrees to ensure streaks of all orientations can be found. The center of the asteroid streak is then randomly paced within 10 pixels of the center of the image. This is to account for the fact that when we deploy the model, some of the streaks input into the model might not be perfectly centered. 

The lengths of each streak are randomly sampled using the following procedure. The distribution of real streak lengths allows us to ensure that the synthetic streaks will roughly match real streaks and the other half of the dataset is to ensure that short and long streaks can still be effectively detected. Although streaks can be longer than 50 pixels, oftentimes they will belong to satellites which move extremely quickly in the sky. Thus, in order to reduce the number of false positives, we set 50 pixels as the maximum limit and focus on shorter streaks.
\begin{enumerate}
    \item 50\% are sampled from a distribution of streak lengths obtained from a set of real asteroid streaks.
    \item 30\% are randomly sampled from a uniform distribution that is between 7 to 12 pixels, which allows us to better target shorter asteroid streaks that are more common than longer ones.
    \item 20\% are randomly sampled from a uniform distribution that is between 30 to 50 pixels, which allows us to also ensure longer ones can still be detected.
\end{enumerate}

The brightnesses of each streak are then randomly sampled in the following steps. This allows us to create a dataset that targets faint streaks while still including some brighter streaks which are easier to find. 
\begin{enumerate}
    \item We first find the approximate minimum visible streak amplitude by multiplying the standard deviation for the region of pixels behind where we want to implant the streak by 3.25. We have experimented with the cutoff and have found that 3.25 is roughly the minimum factor needed for the streak to be visible by a human viewing the image, otherwise the streak is too dim and is hard to tell apart from noise. 
    \item This minimum amplitude is then multiplied by a randomly chosen ``factor". For 80\% of streaks, this factor is randomly selected from the right half of a normal distribution (so we only get numbers above the mean) with mean 1 and standard deviation 0.5. For 20\% of streaks, this factor is randomly selected uniformly from 1.75 to 3.5. 
    \item For shorter asteroids ($<$12 pixels), we increase the factor by 30\% since otherwise they will not be visible due to the small area that they cover.
\end{enumerate}

The widths of each streak are sampled from a distribution of streak widths obtained from a set of real streaks. Generally, the widths of streaks do not vary immensely, as there usually are not any extremely thin or wide streaks, and thus we do not use any other special sampling besides the distribution of widths of real streaks.

\begin{figure}
\centering
\includegraphics[width=0.125\textwidth]{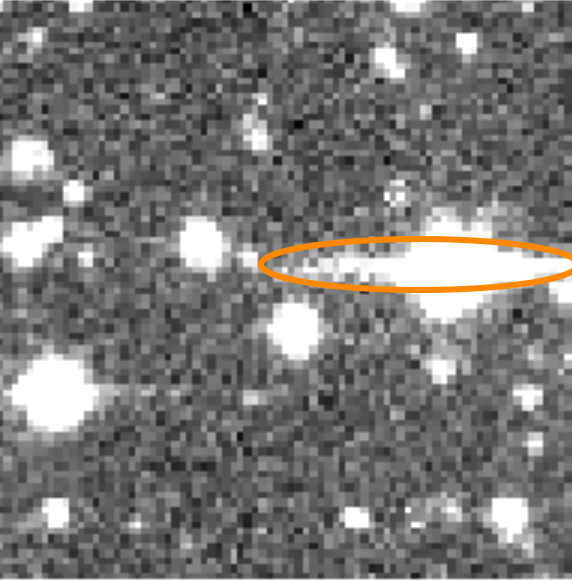}
\caption{Example of a diffraction spike. By not generating streaks which are on top of bright stars, the number of diffraction spike detections drops significantly.}
\label{fig:diffspike}
\end{figure}

To prevent detections of diffraction spikes (see Figure \ref{fig:diffspike}), asteroid streaks are prevented from being inserted on top of a large area which contains pixels 5$\sigma$ above the background. This prevents the model from recognizing streak-like objects which overlap extremely bright stars, which are largely diffraction spike artifacts produced by extremely saturated stars. The introduction of this change to the dataset has resulted in a significant drop in the number of diffraction spike detections. 

\begin{figure}
\centering
\includegraphics[width=0.48\textwidth]{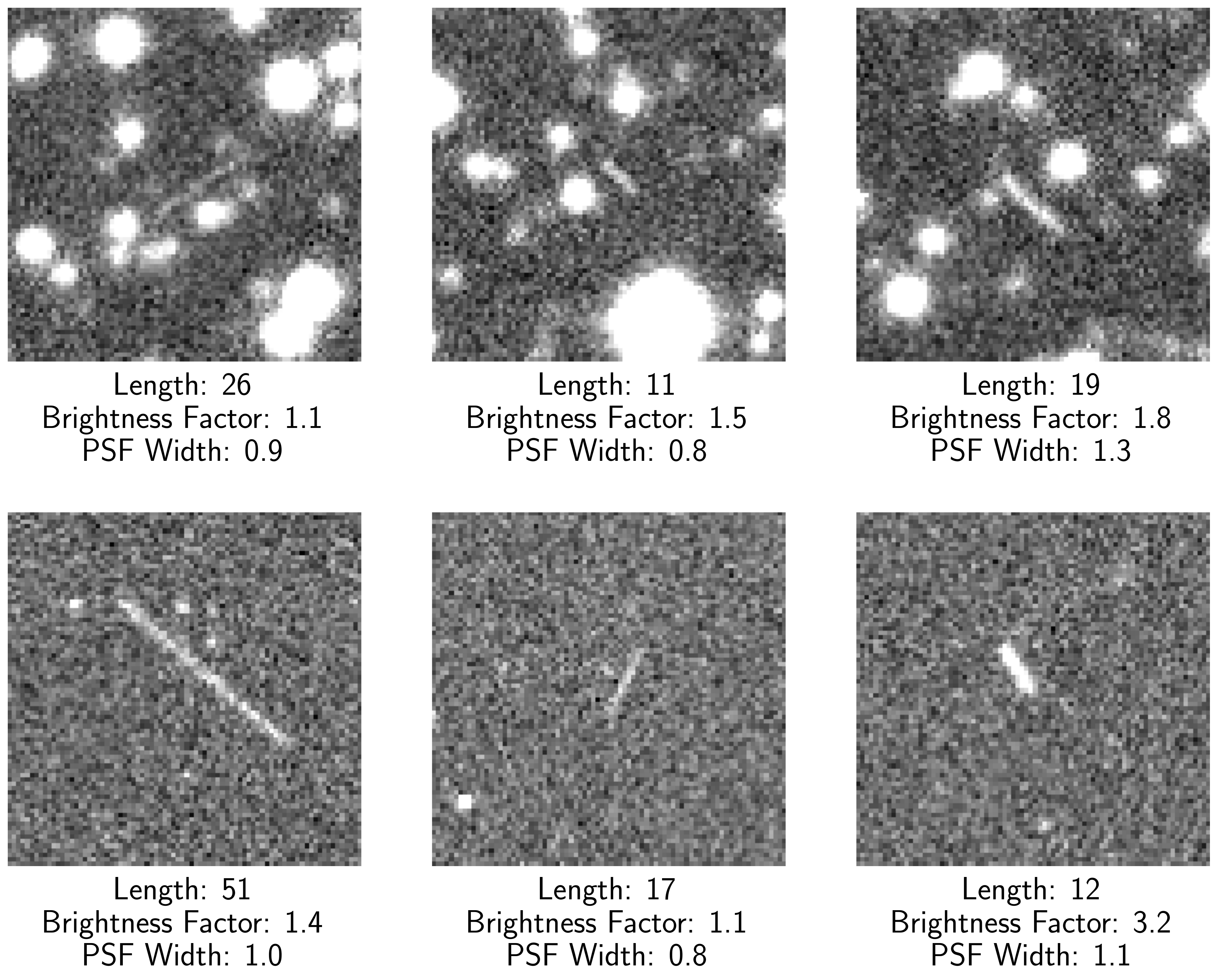}
\caption{Examples of synthetic streaks and the parameters used to create them. Notice when the brightness factor is close to 1 (top left), the streak is extremely faint. The PSF width parameter corresponds to roughly half of the overall width of the streak, but if the streak is very bright it may have a larger width than expected.}
\label{fig:exstreaks}
\end{figure}

Overall, we generate roughly 400,000 streaks (see Figure \ref{fig:exstreaks}) to train our machine learning model. The extremely large size of this dataset ensures that the model is able to be exposed to a large variety of streaks, especially faint ones, and also prevents the model from overfitting on the data, which is where the model simply memorizes each data sample rather than learning how to extrapolate the results to unseen data.

In addition, to ensure that the CNN will not simply recognize streak-like objects in the science image, but will also make sure that they are a transient object not present in the reference image, we generate $\sim$100,000 negative samples consisting of a science and reference pair with the same streak inserted into both pictures. This prevents the CNN from falsely detecting streak-like objects such as galaxies which are not transients and will appear in both the science and reference images.

\subsubsection{Difference Image Preprocessing}

In order to reduce the number of regions the machine learning model has to look at, an initial preprocessing step is used to find potential transients in the difference images which are then fed into the machine learning model. These transients include asteroid streaks but will largely consist of false positives like artifacts, which is why we need a CNN that will take in these transients to determine if they are asteroid streaks. This preprocessing lets us avoid having to search every region of every image, thus reducing the processing time required, and works as follows:
\begin{enumerate}
\item The difference image’s background and background standard deviation is computed using the sep library \citep{Barbary2016}.
\item A mask containing only pixels 1.3 standard deviations above the background is created, which allows us to find bright pixels.
\item Contiguous regions with at least 15 pixels and a fullness greater than 0.5, are kept while other regions are rejected. The fullness is computed by dividing the number of pixels in the region by the number of pixels of the convex hull of the region, which allows us to filter out badly subtracted stars which usually have empty regions within them. We also reject any regions with a fullness exactly equal to 1, since those are usually artifacts shaped like perfect rectangles. 
\item The pixel positions of the centers of each region are stored and then projected onto the respective reference images (this step is to align the pixel positions of the transients on the difference image to the reference image since they will not be perfectly aligned). These projected positions are the output of the pipeline.
\end{enumerate}

\begin{figure}
    \begin{center}
        \includegraphics[width=0.2\textwidth]{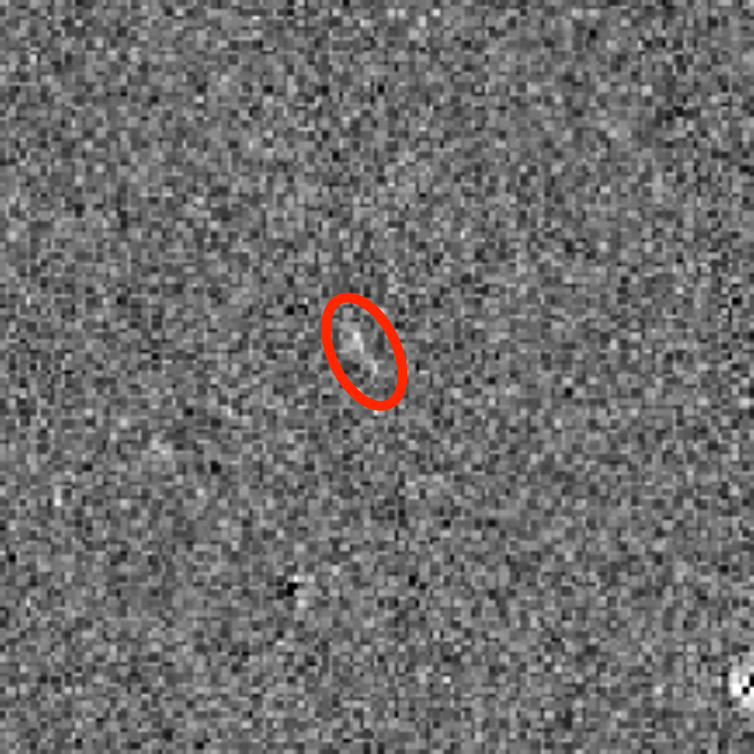}
        \includegraphics[width=0.2\textwidth]{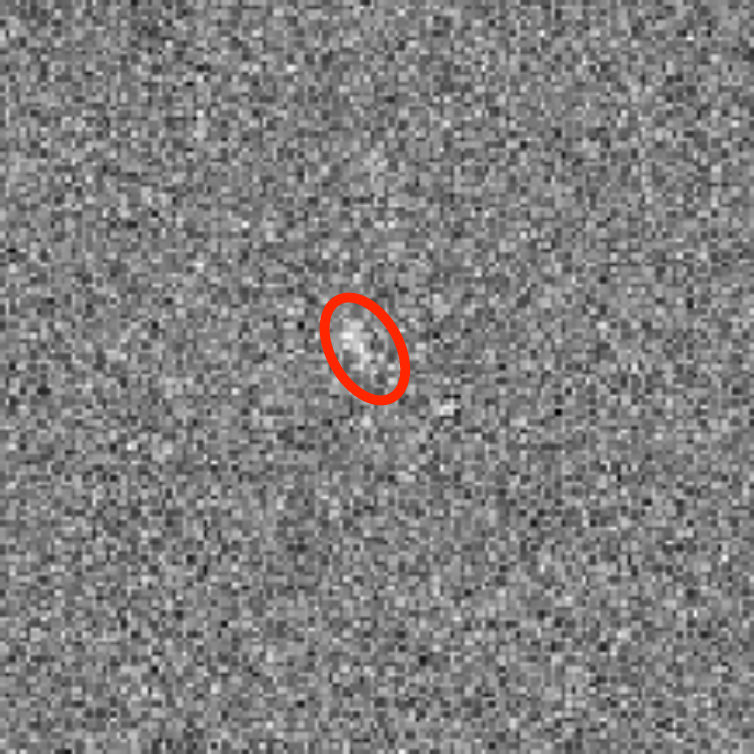}
        \includegraphics[width=0.2\textwidth]{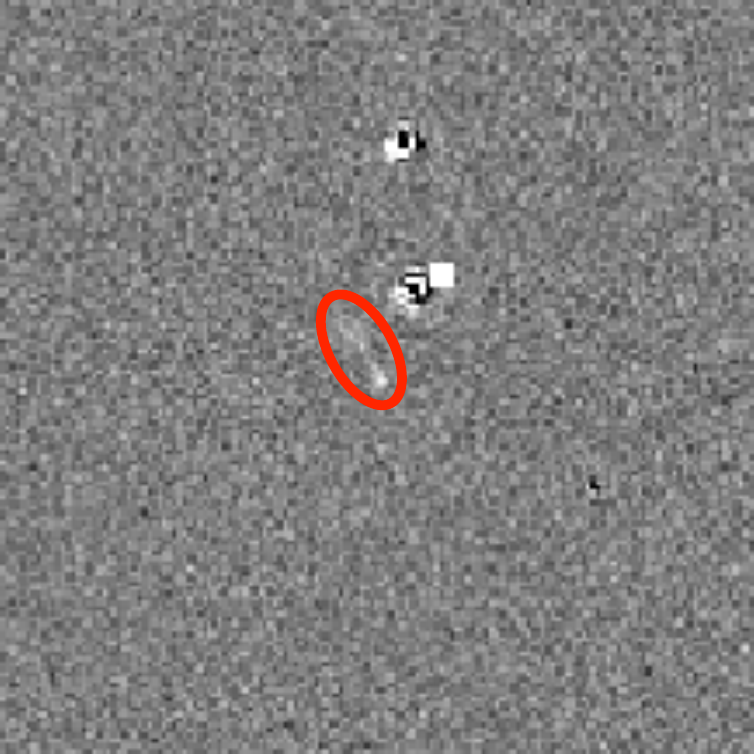}
        \includegraphics[width=0.2\textwidth]{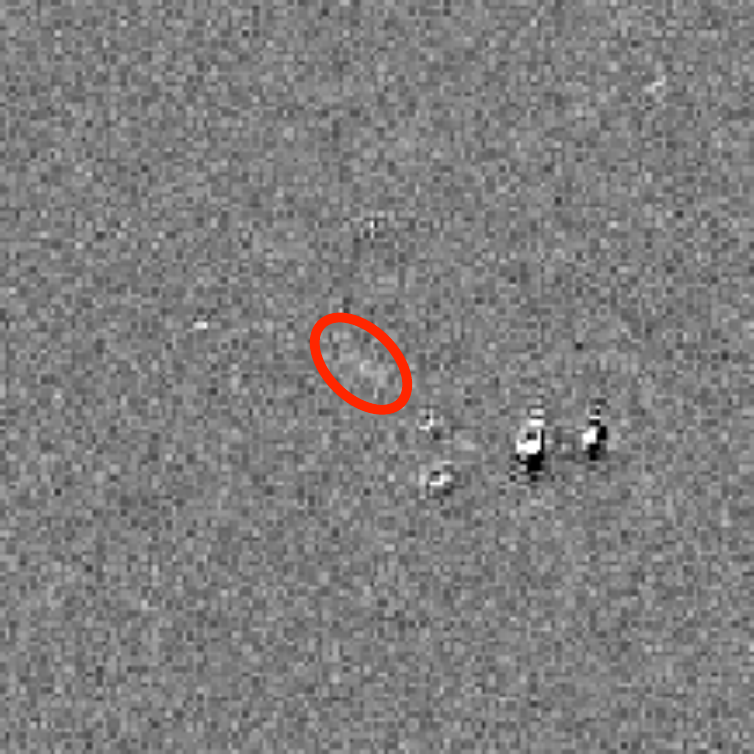}
    \end{center}
    \caption{Examples of very faint real asteroid streaks detected by the difference image preprocessing algorithm.}
    \label{fig:detstreaks}
\end{figure}

\begin{figure}
    \begin{center}
        \includegraphics[width=0.2\textwidth]{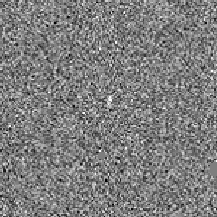}
        \includegraphics[width=0.2\textwidth]{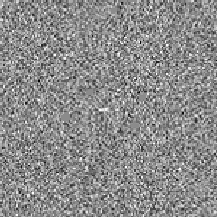}
        \includegraphics[width=0.2\textwidth]{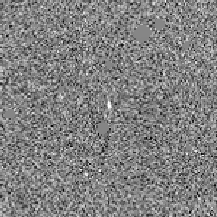}
        \includegraphics[width=0.2\textwidth]{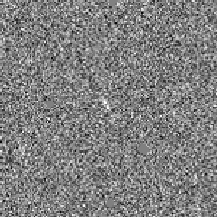}
    \end{center}
    \caption{Examples of real asteroid streaks missed by the difference image preprocessing algorithm. All of the missed detections consist of extremely short and faint streaks.}
    \label{fig:failedstreaks}
\end{figure}

This preprocessing method has significant advantages over ZStreak, ZTF's streak processing pipeline, as it can detect vertical and horizontal streaks because it does not use the Pearson correlation coefficient for filtering and has much less restrictions on the properties of streaks. For instance, ZStreak employs a threshold of 1.5 $\sigma$ while we only use a 1.3$\sigma$ threshold, allowing for dimmer objects to be found \citep{deepstreaksduev}. Our method is also extremely fast due to its simplicity and the fact that it does not use any complex algorithms like streak fitting. In addition, the thresholds imposed on their properties of the transients are very relaxed, allowing for more complete detections of streaks. The thresholds were tuned so that roughly 98\% of streaks in our real streaks dataset would be detected (only 9 out of 402 streaks failed). Examples of streaks which are detected by the preprocessing step can be found in Figure \ref{fig:detstreaks} and ones which are not can be found in Figure \ref{fig:failedstreaks}. The ones that were not detected were all extremely short and faint, barely visible to the human eye.

The vast majority of the detections produced by this pipeline are not real asteroid streaks. Because of this, we run this pipeline on random difference images, then extract the respective science and reference image regions corresponding to the positions returned to create negative samples for our dataset, totaling to around 400,000 negative samples to be used for our convolutional neural network. These negative samples overwhelmingly consist of artifacts and random noise. Additionally, we use the \textit{astcheck} program to ensure that these negative samples do not contain any known asteroids, greatly reducing the number of incorrect training samples as it is extremely unlikely for there to be an undiscovered asteroid streak in the negative samples\footnote{\url{https://www.projectpluto.com/astcheck.htm}}.

\subsection{Neural Network Model}\label{sec:nnmodel}

The classifier that we use in this paper to find asteroid streaks is a convolutional neural network (CNN), specifically a modified version of EfficientNet-B1 \citep{effnettanle}. CNNs have had much success in being able to differentiate between different classes of images and achieve state-of-the-art performance on classification tasks like ImageNet, which involves classification of images into object categories like birds, cars, and screwdrivers \citep{russakovsky_imagenet_2015}. CNNs accomplish this through combination of many different ``layers" which are trained to recognize images using training data and gradient descent.

EfficientNet is a family of CNN models that currently has one of the highest accuracy to performance ratios, which is important as we want a model that runs quickly and has high accuracy at the same time. EfficientNet is able to achieve its high efficiency by computationally searching for the best model configuration and method for scaling up the size of the CNN \citep{effnettanle}. 

We chose to use EfficientNet-B1 (the 2nd smallest version of the model) in order to allow for higher accuracy than the B0 model (the smallest version) while still allowing the CNN to be run quickly on most GPUs. Because the EfficientNet model is intended to be used on RGB images from the ImageNet dataset, we modify the inputs to contain two 80 by 80 images -- the science and reference cutouts -- and the output to be a probability from 0 to 1 -- representing the probability the image contains an asteroid streak. This is done using an EfficientNet-B1 backbone and then modifying the final layers to use global pooling and fully connected layers to process the feature vector output by EfficientNet (see Table \ref{tab:nnarch}). 

\begin{figure} 
\centering
\includegraphics[width=0.48\textwidth]{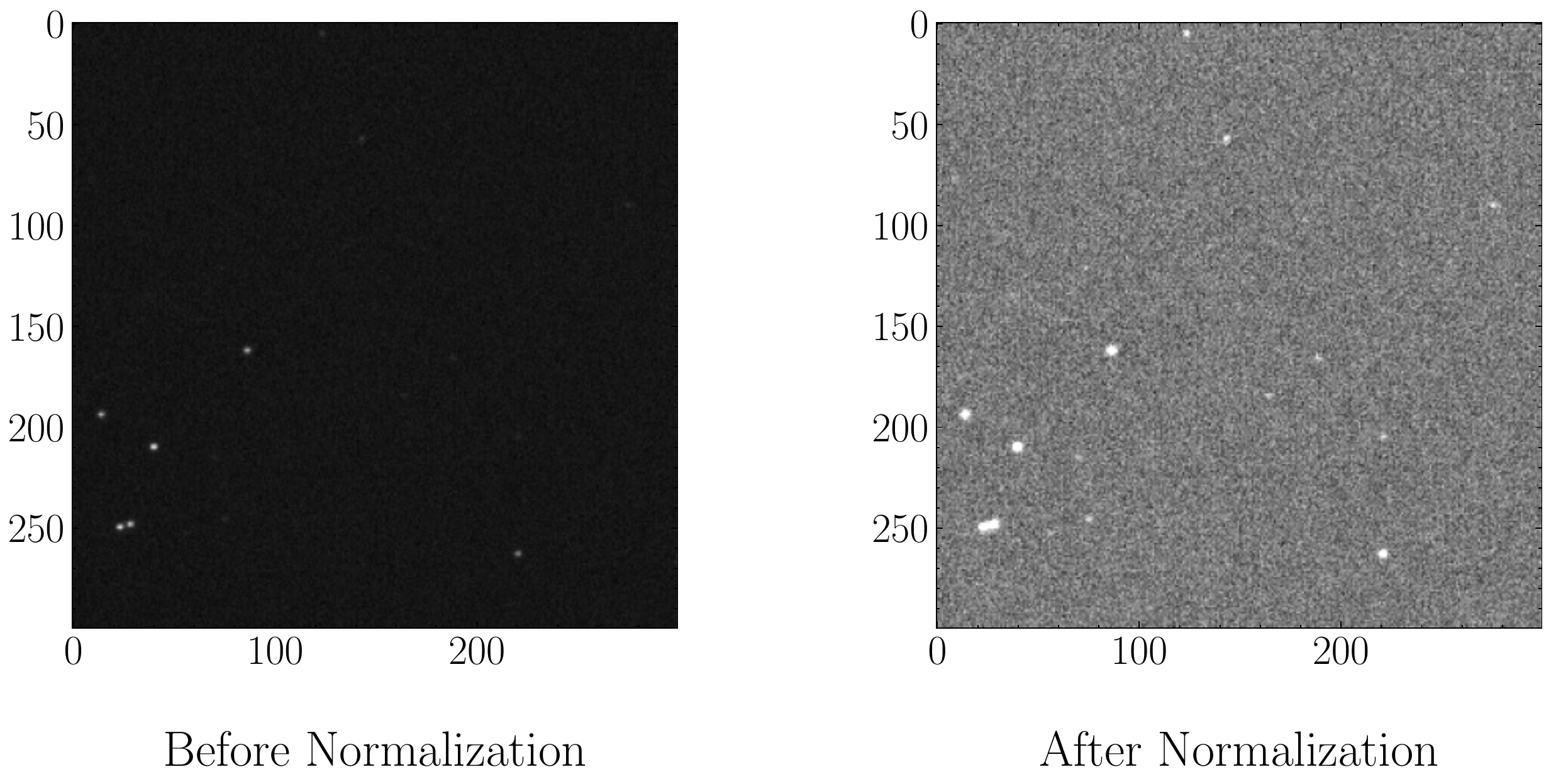}
\caption{Example of using the normalization algorithm to normalize an image region. Notice how in the left image the bright stars on the bottom left dominate the image, making the rest of the image almost invisible. When we normalize the image (as shown on the right), those bright pixel values are clipped, and thus we can better see the rest of the image.}
\label{fig:normimg}
\end{figure}

The raw image data can not be passed directly into the machine learning model as the pixel values can vary widely due to extremely bright stars. If we input these numbers directly into a neural network, we may run into overflow errors. Moreover, extremely bright stars will dominate images and make dim streaks almost invisible in comparison (see Figure \ref{fig:normimg}). However, the full range of brightnesses is not necessary, especially since this paper focuses on fainter asteroids. 

To normalize both the science and reference images, the image background intensity and background RMS is computed using the \textit{sep} library, which is based upon the popular Source Extractor library \citep{Barbary2016,bertin_sextractor}. This allows us to standardize each pixel in the image by computing the difference between the pixel and the background intensity and then dividing that difference by the standard deviation. Pixel values below $-5$ and above $5$ are clipped to $-5$ and $5$, respectively, since extremely low and high pixel values do not provide much useful information, allowing us to focus on dimmer pixels which may belong to streaks. This gives us the following equation for normalizing each pixel $p$:
\begin{equation}\label{eq:norm}
    p_\text{norm} = \max\left(-5, \min\left(5, \frac{p - \mu_\text{bkg}}{\sigma_\text{bkg}}\right)\right)
\end{equation}

To reduce overfitting, where the CNN ``memorizes" the training data rather than learning how to extrapolate to unseen samples, we apply a dropout rate of 0.2, which causes 20\% of neurons in the CNN to be randomly omitted during each iteration of training. Dropout has been shown to be effective in reducing overfitting by making it harder for neural networks to reliably memorize training data \citep{dropout_srivastava_et_al}.

This model was implemented using the Keras \citep{chollet2015keras} and EfficientNet Python libraries\footnote{\url{https://github.com/qubvel/efficientnet}}. To train the CNN, we used the binary cross-entropy loss function, a smaller batch size of size 32 to further reduce overfitting \citep{masters_luschi_batch} and the Adam optimizer \citep{kimga_ba_adam}. We used an NVIDIA 2080 TI GPU to increase training speed.

\begin{table}
\centering
\begin{tabular}{c c c c}
\hline
Layer & Input Shape & Output Shape & \# Params \\ \hline
EfficientNet-B1 Backbone & (80, 80, 2) & (3, 3, 1280) & 6574944\\
Global Avg Pooling & (3, 3, 1280) & 1280 & 0 \\
Fully Connected & 1280 & 256 & 327936 \\
Fully Connected & 256 & 1 & 257
\end{tabular}
\caption{\label{tab:nnarch} The modified EfficientNet-B1 architecture that we use.}
\end{table}

\subsection{False Positive Reduction} \label{sec:artifacts}

To reduce the number of false positives the model outputs, artifacts which may be detected as a false positive must be removed. The three types of false positives prevalent in ZTF data are cosmic rays, ghosts, and crosstalk. 

\begin{figure}
\centering
\begin{minipage}{0.2\textwidth}
    \centering
    \includegraphics[width=0.95\textwidth]{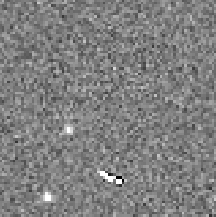}
\end{minipage}
\begin{minipage}{0.2\textwidth}
    \centering
    \includegraphics[width=0.95\textwidth]{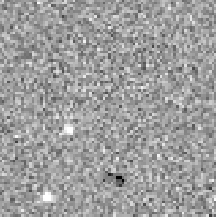}
\end{minipage}
\caption{Example of a cosmic ray artifact (left image), which is then removed using AstroSCRAPPY (right image). Notice how the cosmic ray has very sharp edges while asteroid streaks have soft edges, which reflects the differing PSFs that can be exploited to detect and remove cosmic rays.}
\label{fig:cosmicrays}
\end{figure}

Cosmic rays are produced when highly energized particles strike CCDs. This produces transient streak-like objects in the images (see Figure \ref{fig:cosmicrays}). One of the weaknesses of our normalization algorithm is that it is hard for the CNN to differentiate asteroid streaks and cosmic rays since cosmic rays are distinct in that they are very bright and their pixels are not distributed according to a standard point spread function, but these nuances are lost due to the clipping of bright pixels. However, there have been many algorithms developed that can effectively exploit the unique appearances of cosmic rays to remove them from images. To remove cosmic rays, we employ the AstroSCRAPPY library, which is based upon the LA Cosmic algorithm \citep{astroscrappy}. The LA Cosmic algorithm uses Laplacian edge detection to differentiate cosmic rays from other sources of light \citep{van_Dokkum_2001}. This allows us to detect cosmic rays in an image, and then mask them out, as is done in Figure \ref{fig:cosmicrays}.

\begin{figure}
\centering
\begin{minipage}{0.45\textwidth}
    \centering
    \includegraphics[width=\textwidth]{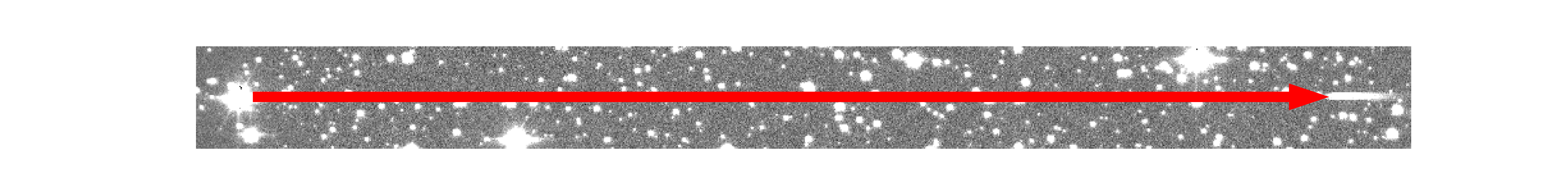}
\end{minipage}
\centering
\begin{minipage}{0.45\textwidth}
    \centering
    \includegraphics[width=\textwidth]{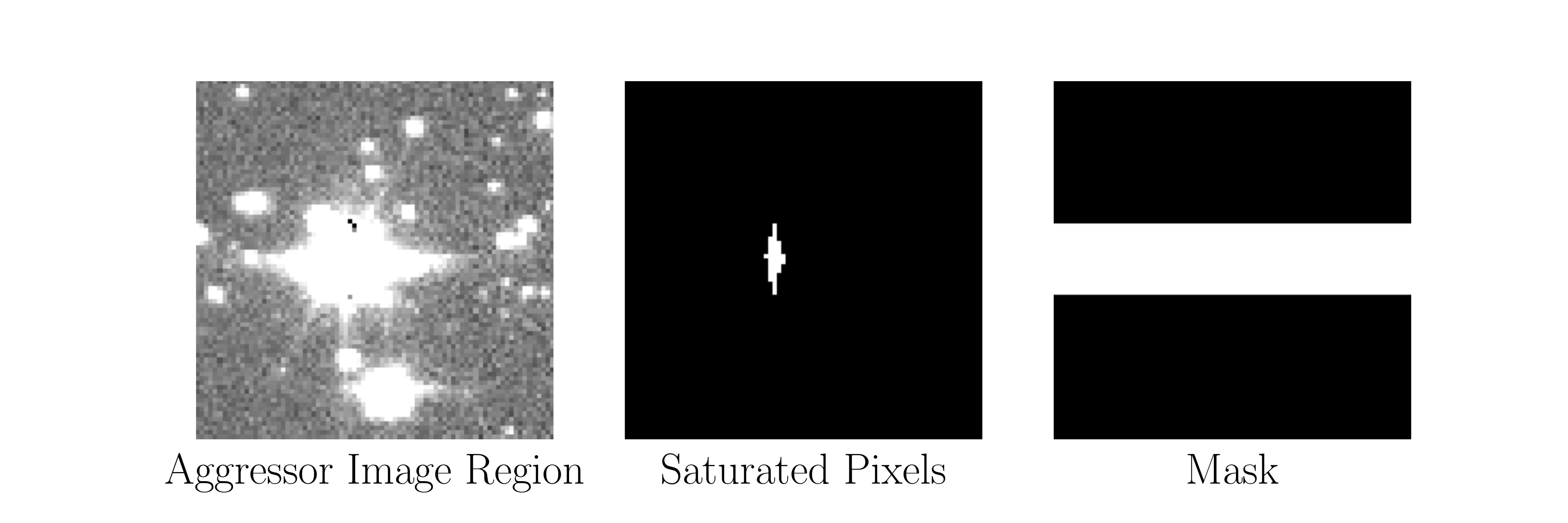}
\end{minipage}
\centering
\begin{minipage}{0.4\textwidth}
    \centering
    \includegraphics[width=\textwidth]{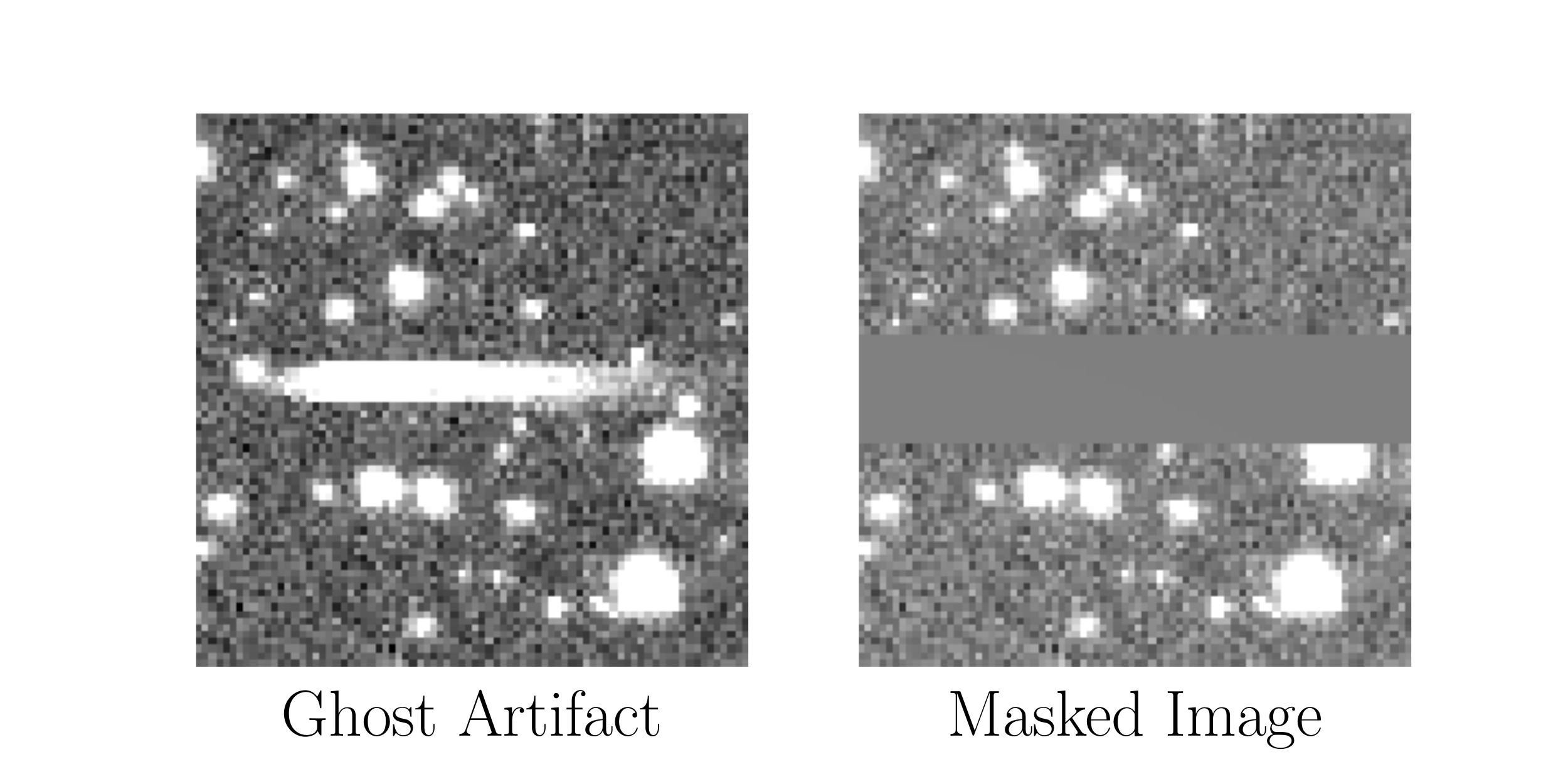}
\end{minipage}
\caption{Example of a ghost artifact, which appears as a horizontal streak-like object on the right of the top image. The top image shows how saturated pixels of a star in one area of an image can cause charge spillage ghosts to appear in another area of the image. The second row of images shows how the saturated pixels can be extracted and converted to a mask. The final row shows how this mask can be applied to remove ghost artifacts.}
\label{fig:ghosts}
\end{figure}

Ghosts are created by charge spillages from saturated pixels in the same image. These ghosts are horizontal streaks located at a set distance in front of or behind a cluster of saturated pixels and are typically up to 50 pixels in width \citep{zwicky_transient_facility_ztf_2020}. To remove these, we mask out pixels which are a certain distance in front of or behind a saturated pixel and the 50 pixels in front and behind it to account for the width (see Figure \ref{fig:ghosts}). Although this sometimes leads to masking of pixels without ghosts, since charge spillage does not always occur, the number of pixels masked in an image is extremely small compared to the total number of pixels in the image (<1\%). 

\begin{figure}
\centering
\begin{minipage}{0.45\textwidth}
    \centering
    \includegraphics[width=\textwidth]{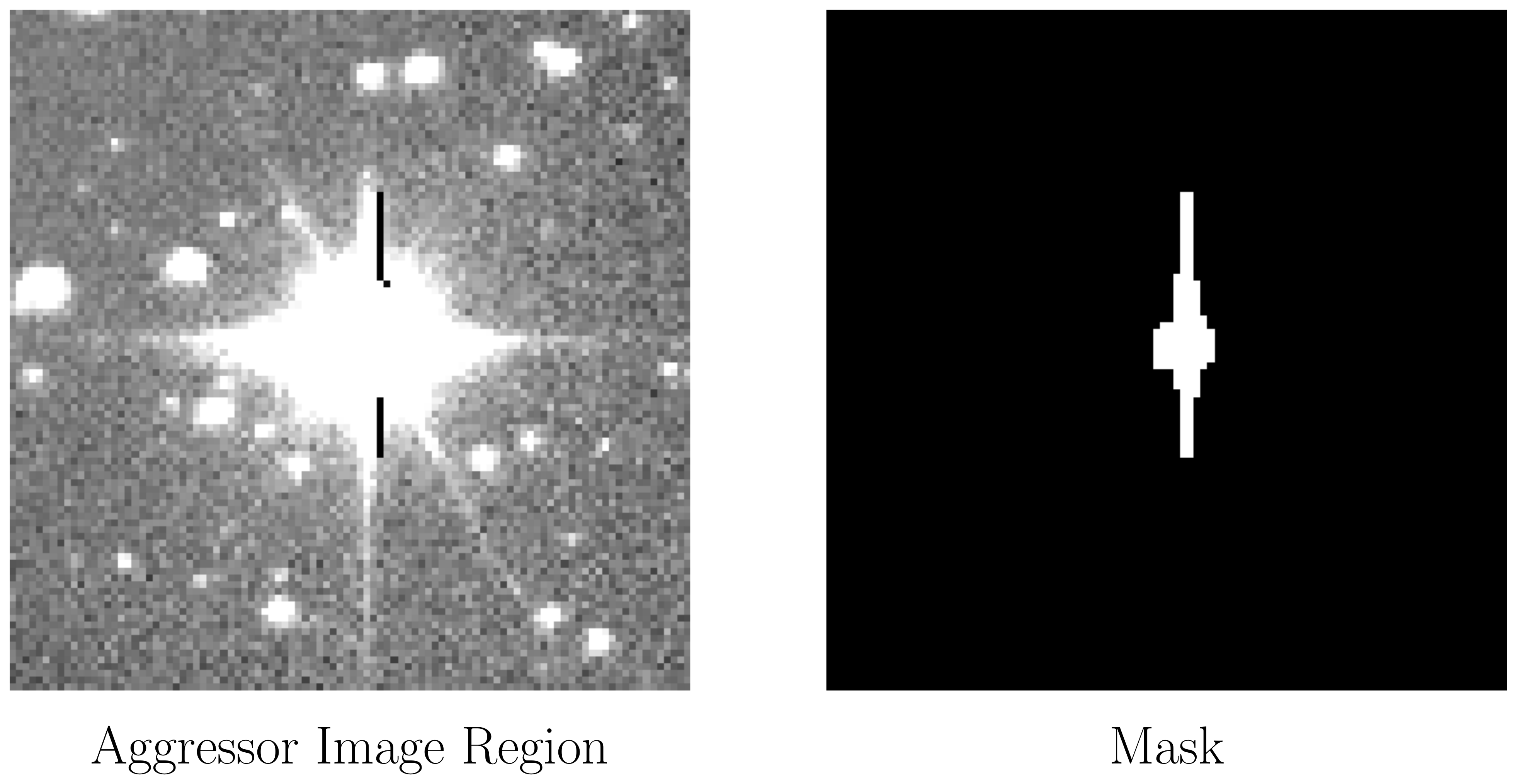}
\end{minipage}
\centering
\begin{minipage}{0.45\textwidth}
    \centering
    \includegraphics[width=\textwidth]{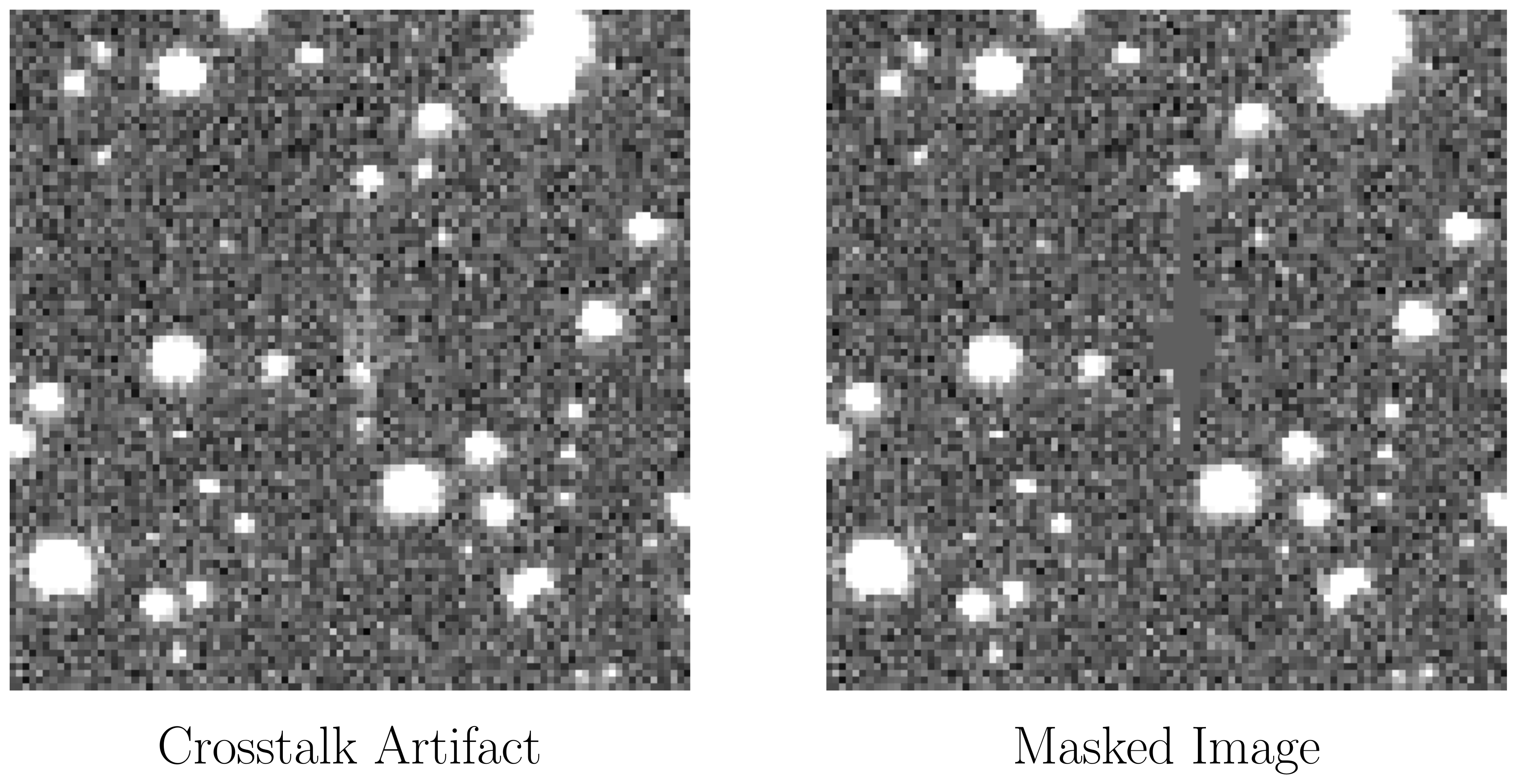}
\end{minipage}
\caption{Example of crosstalk. The aggressor pixels on the top left image (quadrant 2), create crosstalk artifacts in the bottom left image (quadrant 1). A mask, as shown by the top right image, is generated from saturated pixels in quadrant 2 and is used to mask out the crosstalk to get the image in the bottom right image. Note that there are two stars on top of the crosstalk on the bottom half of the artifact which may make it appear as if the artifact was not completely masked out.}
\label{fig:crosstalk}
\end{figure}

Crosstalk is an artifact produced when a strong signal in one CCD quadrant (channel) is electronically replicated onto another quadrant. To remove such artifacts, the pixel locations of strong signals (e.g. saturated pixels) are masked in both quadrants affected by crosstalk (see Figure \ref{fig:crosstalk}). We observed that the CCD pairs affected by bi-directional crosstalk are quadrants 1-2 and 3-4.

\subsection{Full Pipeline}

The components used for training the neural network are combined together to create a pipeline for detecting asteroids. To process one night of data, we use the following procedure on every single triplet of the corresponding science, reference, and difference images to search for asteroid streaks. The result of this is a large set of potential asteroid streak candidates.
\begin{enumerate}
    \item The difference image preprocessing pipeline is used to extract transient objects, which will include asteroid streaks but mostly artifacts and other non-streak transients. The positions of these transients are projected from the difference image onto the reference image. 
    \item The science image is aligned to the reference image using the SWarp program \citep{bertinswarp}. This is to account for differences in rotation and position of the science and reference images. The positions obtained from Step 1 are used to extract 80 by 80 pixel crops from the aligned science and reference images, where the center of the crop corresponds to the center of the transients. 
    \item All the crops are normalized using the normalization process described in Section \ref{sec:nnmodel} and Equation \ref{eq:norm}. These normalized images are then inputted into the trained convolutional neural network, which outputs probabilities for each science-reference pair.
    \item These probabilities are thresholded using a threshold of 0.85, so that probabilities above 0.85 are considered to be positive detections and ones below 0.85 are considered to be negative detections. This gives us a set of images which the CNN believes contain asteroid streaks.
    \item Cosmic rays, crosstalk, and ghosts are removed from the original un-normalized images of the positive detections if present and the cleaned images are re-normalized and re-inputted into the CNN. This allows us to reduce computation time by only running the false positive reduction on the small subset of positive detections from the CNN.

\end{enumerate}
Each streak candidate is then visually checked and false positives are removed. Detections with the same orientation and length along with a corresponding motion rate and movement direction are linked together. The \textit{astcheck}\footnote{\url{https://www.projectpluto.com/astcheck.htm}} program is then used to check if the detection corresponds to a previously discovered asteroid and the \textit{sat\_id}\footnote{\url{https://www.projectpluto.com/sat_id.htm}} program is used to ensure it does not correspond to a known artificial satellite. 

\section{Results} \label{sec:results}

\subsection{CNN Training Results}

The validation dataset that we use consists of a subset of our dataset which is set aside and not trained on, so that we can evaluate the performance of our model on unseen data. We generate additional positive and negative samples so that in total, 30\% of our data is used for the validation set, which is abnormally high for most machine learning research. However since our false positive rate is extremely low ($\sim$0.02\%), we must ensure that we have enough data to obtain an accurate estimate of the false positive rate. 

\begin{figure}
\centering
\includegraphics[width=0.45\textwidth]{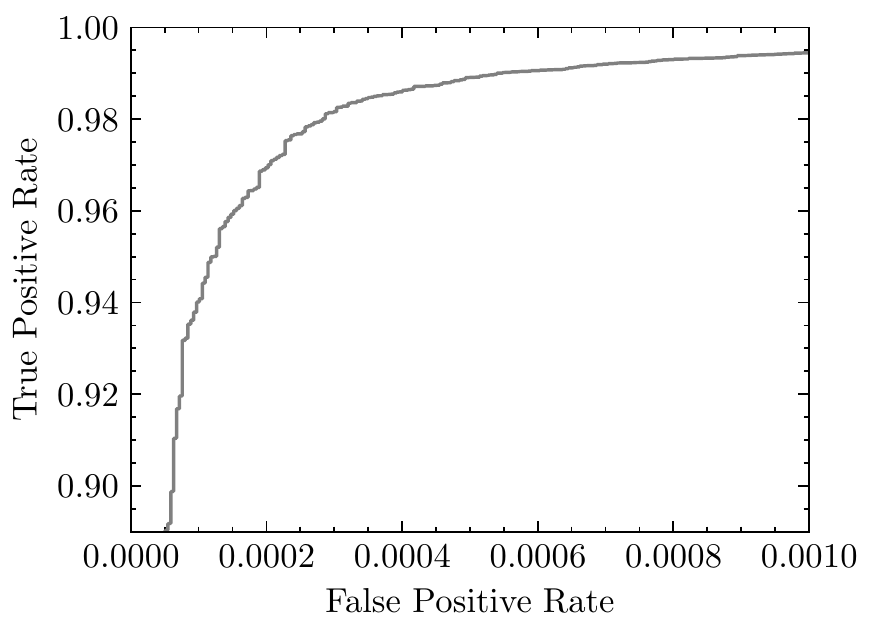}
\caption{The receiving operator characteristic plot which shows the true positive rate required to have a certain false positive rate. Overall, the false positive rate is extremely small for true positive rates less than 98\%.}
\label{fig:roc}
\end{figure}

The final performance of the CNN on the validation dataset can be evaluated on the metrics in Table \ref{tab:metrics} and the receiving operator characteristic (ROC) curve which plots the true positive rate against the corresponding false positive rate (see Figure \ref{fig:roc}). Overall, the accuracy for the validation dataset is extremely high at 98.7\% while still maintaining an extremely low false positive rate of just 0.02\%. This low false positive rate is critical to reducing the number of images outputted by the CNN per night since the vast majority of images do not contain asteroid streaks.

\begin{table}
\centering
\begin{tabular}{c c c c}\hline
    \textbf{Val. Acc} & \textbf{False Positive Rate} & \textbf{True Positive Rate} & \textbf{ROC AUC}\\
    \hline
    98.7\% & 0.02\% & 97.0\% & 99.99\%
\end{tabular}
\caption{Metrics showing the performance of the convolutional neural network. The ROC AUC is the area under the curve of the receiving operator characteristic which plots the true positive rate against the false positive rate for different thresholds (see Figure \ref{fig:roc}).}
\label{tab:metrics}
\end{table}

\subsection{CNN Performance on ZTF's Reported Streaks}\label{sec:cnnperformance}
We collected a set of real streaks from ZTF's submissions to the Minor Planet Center. The submissions falling within the date range of ZTF's publicly available data were then kept, and cutouts of the corresponding streaks are extracted. This dataset of real streaks differs from the one used to generate the distributions of our simulated data since it only uses streak detections which ZTF has verified and reported to the MPC. This allows us to directly compare our model's performance on streaks which ZTF's DeepStreaks model \citep{deepstreaksduev} has successfully detected. We incorporate data from detections of confirmed asteroids and unconfirmed asteroids (which are objects with a short observational arc located in the MPC's Isolated Tracklet File). Since this research focuses on shorter streaks, we remove streaks longer than 50 pixels in length. 

\begin{figure*}
    \centering
    \includegraphics[width=0.55\textwidth]{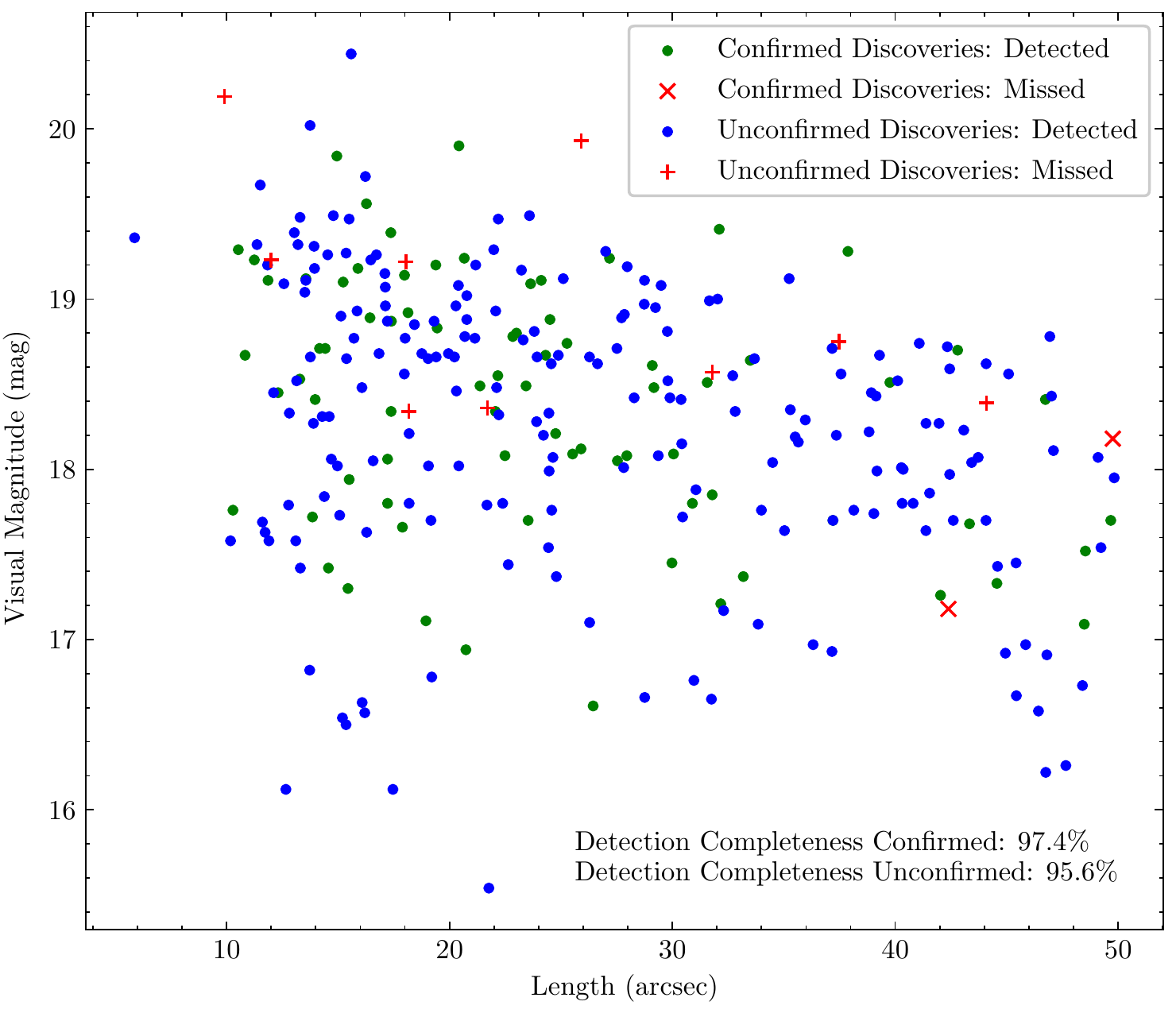}
    \caption{Visual magnitude and length of streaks \textbf{discovered} by ZTF in addition to whether our CNN was able to detect them. These streaks are divided into confirmed and unconfirmed (which tend to be fainter and thus more difficult to obtain a longer observational arc).}
    \label{fig:detcompltdets}
\end{figure*}

\begin{figure*}
    \centering
    
    \begin{minipage}{0.45\textwidth}
        \centering
        \includegraphics[width=0.6\textwidth]{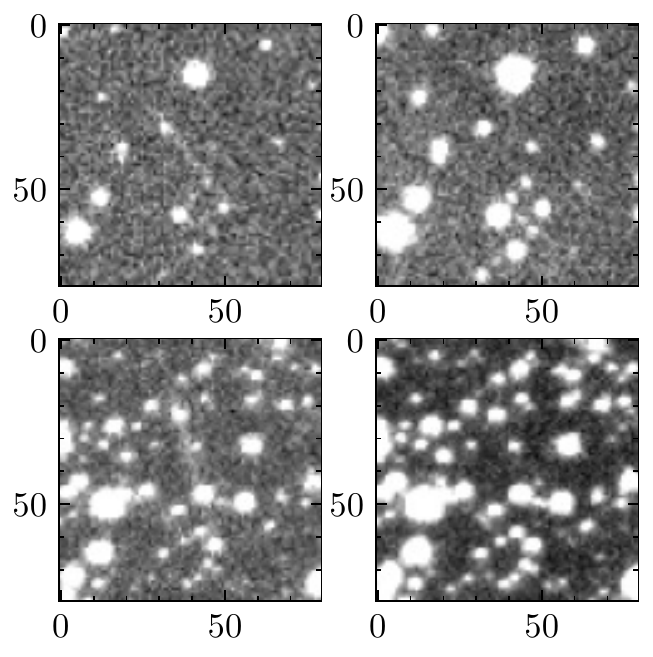}
        \caption{All images of \textbf{confirmed} streaks discovered by ZTF that were incorrectly classified by our model. Left column consists of science images and right column consists of reference images.}
        \label{fig:failedconf}
    \end{minipage}%
    \hspace{20pt}
    \begin{minipage}{0.45\textwidth}
        \centering
        \includegraphics[width=0.55\textwidth]{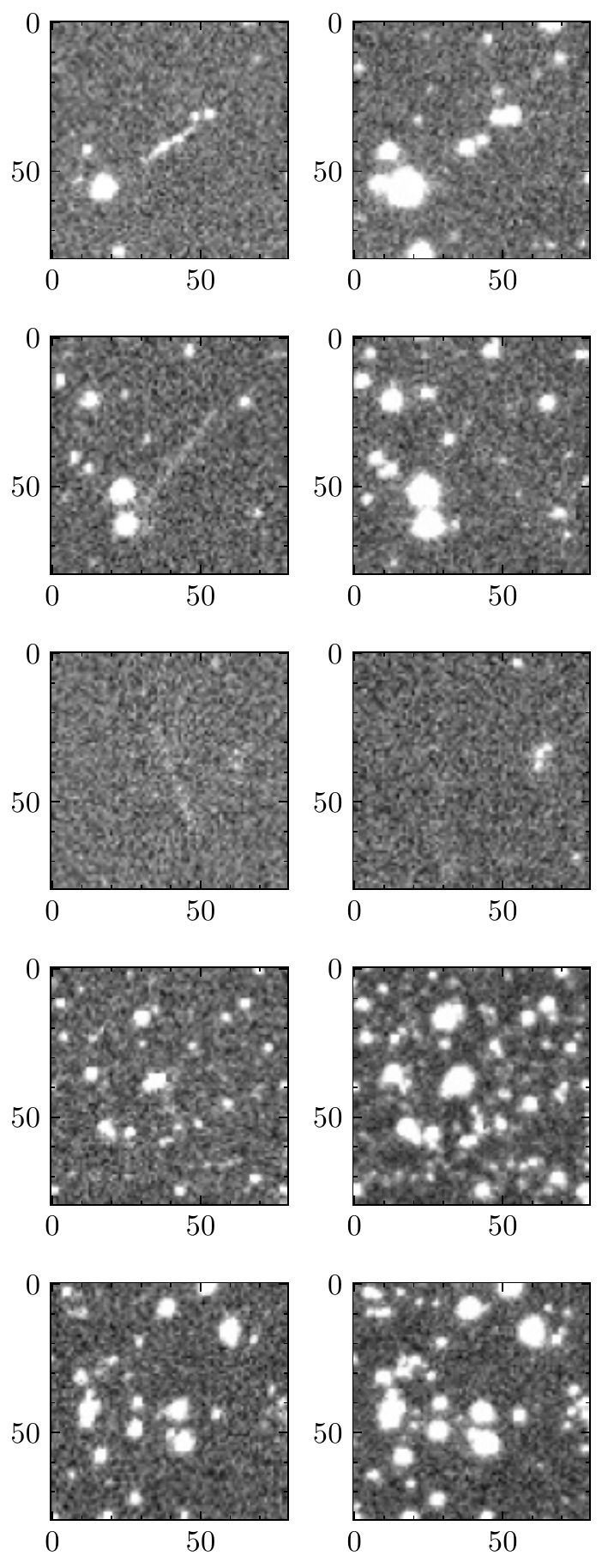}
        \caption{A random sample of five images of \textbf{unconfirmed} streaks discovered by ZTF that were incorrectly classified by our model. Left column consists of science images and right column consists of reference images. The final streak is heavily occluded by a star.}
        \label{fig:failedunconf}
    \end{minipage}
    
\end{figure*}

After running our CNN on the original images that contain these previously detected streaks, we plotted the visual magnitude and motion rates (lengths of streaks) of the correct and incorrect classifications (see Figure \ref{fig:detcompltdets}). Overall, our detection completeness is very high, roughly matching the true positive rate for simulated data. The detection completeness for confirmed asteroids is 97.4\% and 95.6\% for unconfirmed asteroids. For simulated data the true positive rate is 97.0\%. We plot the missed detections in Figures \ref{fig:failedconf} and \ref{fig:failedunconf}. 

\subsection{Discoveries Overview}

Our pipeline was run on the following nights of data from the ZTF public data releases: 2019/06/05-08. The DeepStreaks algorithm began making discoveries in November of 2018, so the data from 2019 allows us to verify that the streaks which we detect were not found by DeepStreaks, thus demonstrating improvement over their algorithm. 

\begin{figure*}
    \centering
    
    \begin{minipage}{0.45\textwidth}
        \centering
        \includegraphics[width=0.95\textwidth]{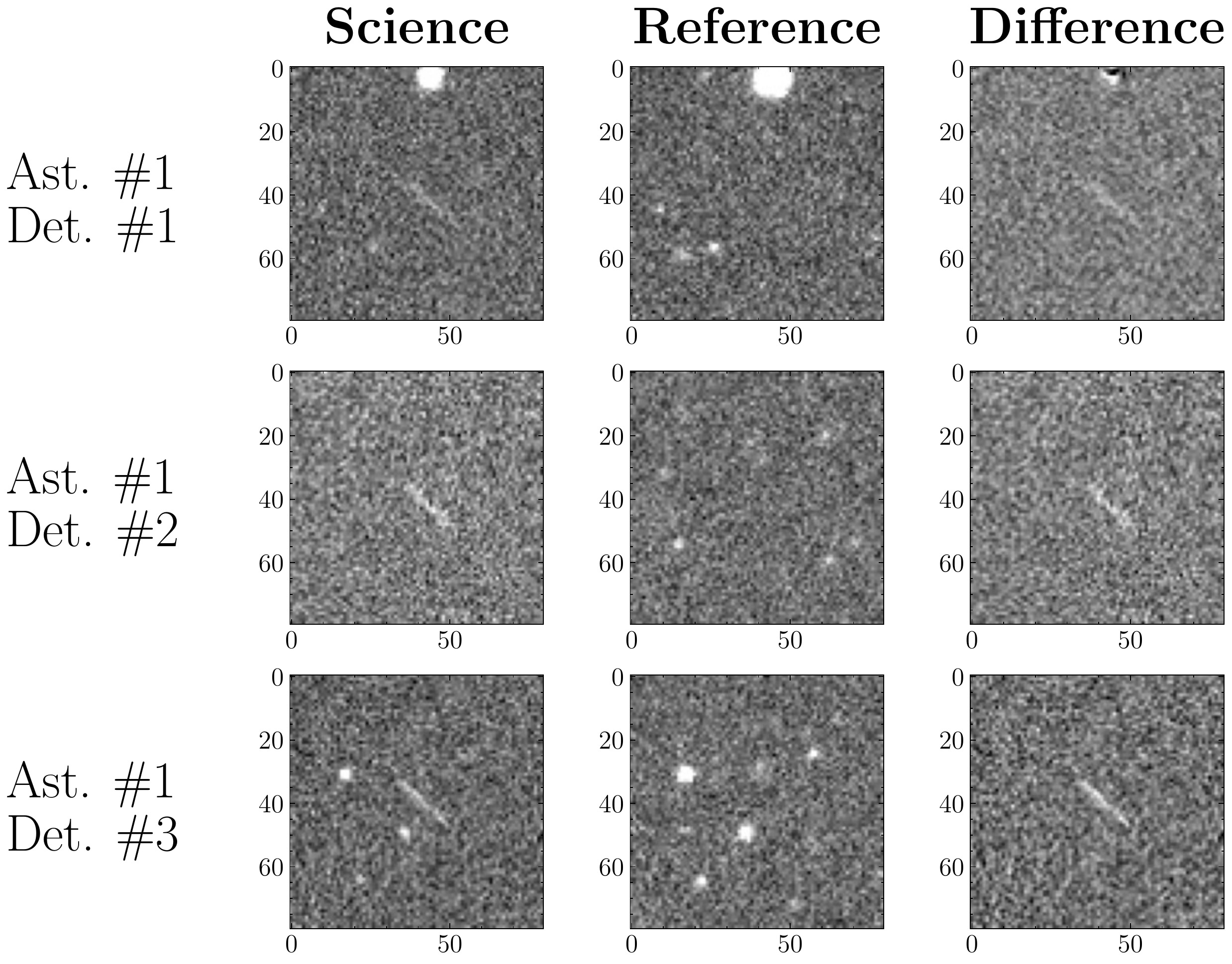}
    \end{minipage}
    \begin{minipage}{0.45\textwidth}
        \centering
        \includegraphics[width=0.95\textwidth]{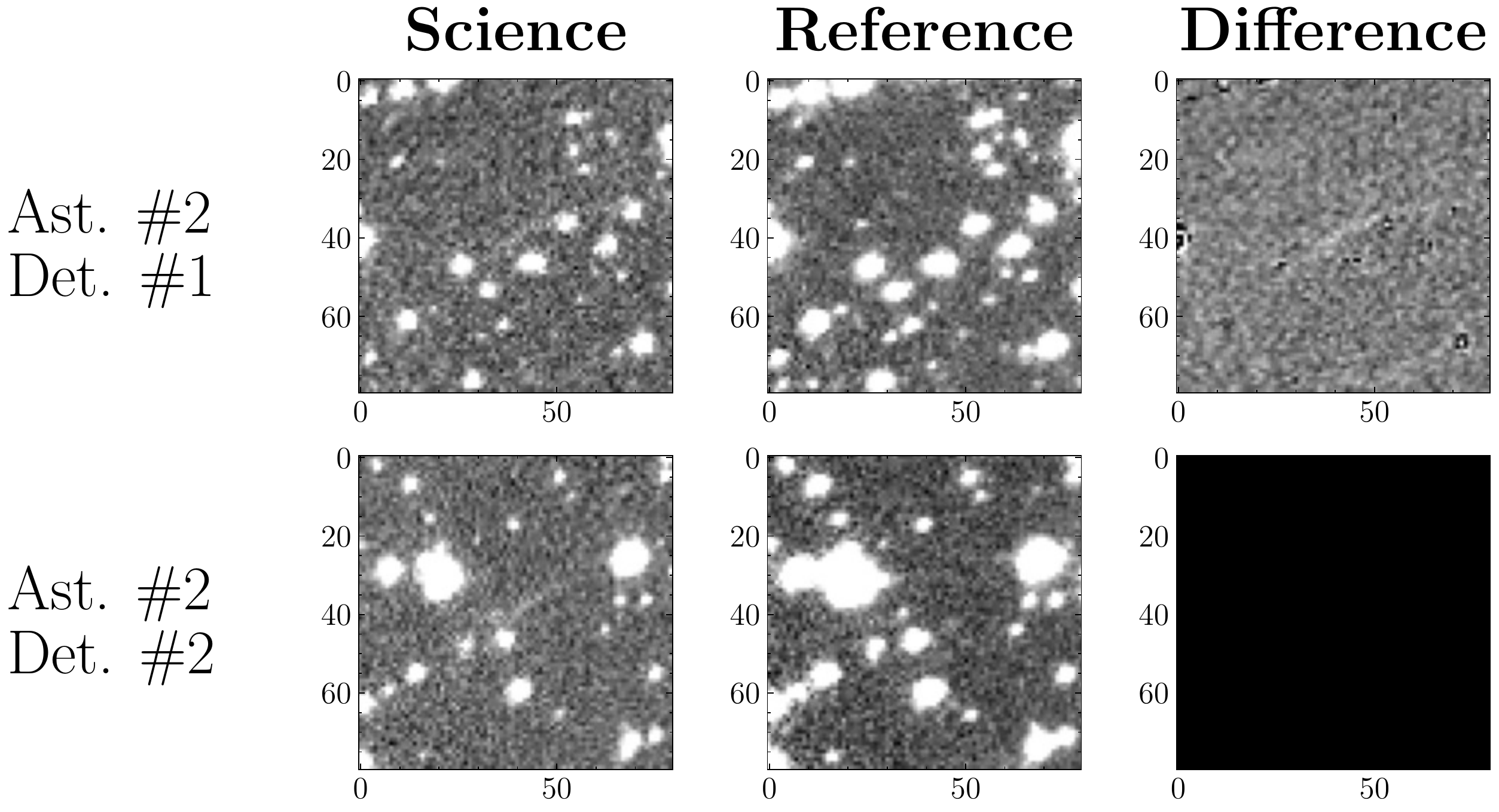} 
    \end{minipage}
    \begin{minipage}{0.45\textwidth}
        \centering
        \includegraphics[width=0.95\textwidth]{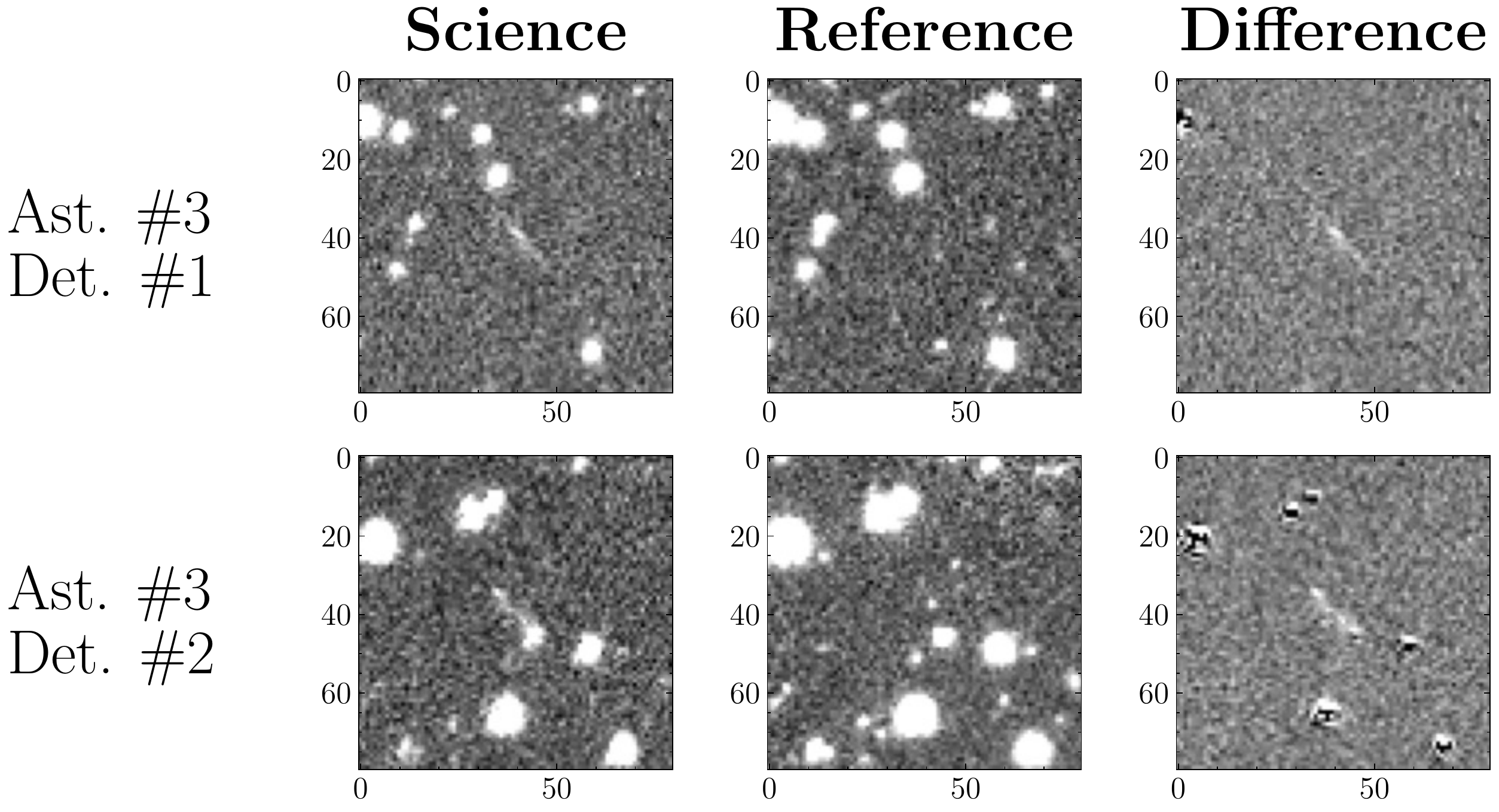}
    \end{minipage}
    \begin{minipage}{0.45\textwidth}
        \centering
        \includegraphics[width=0.95\textwidth]{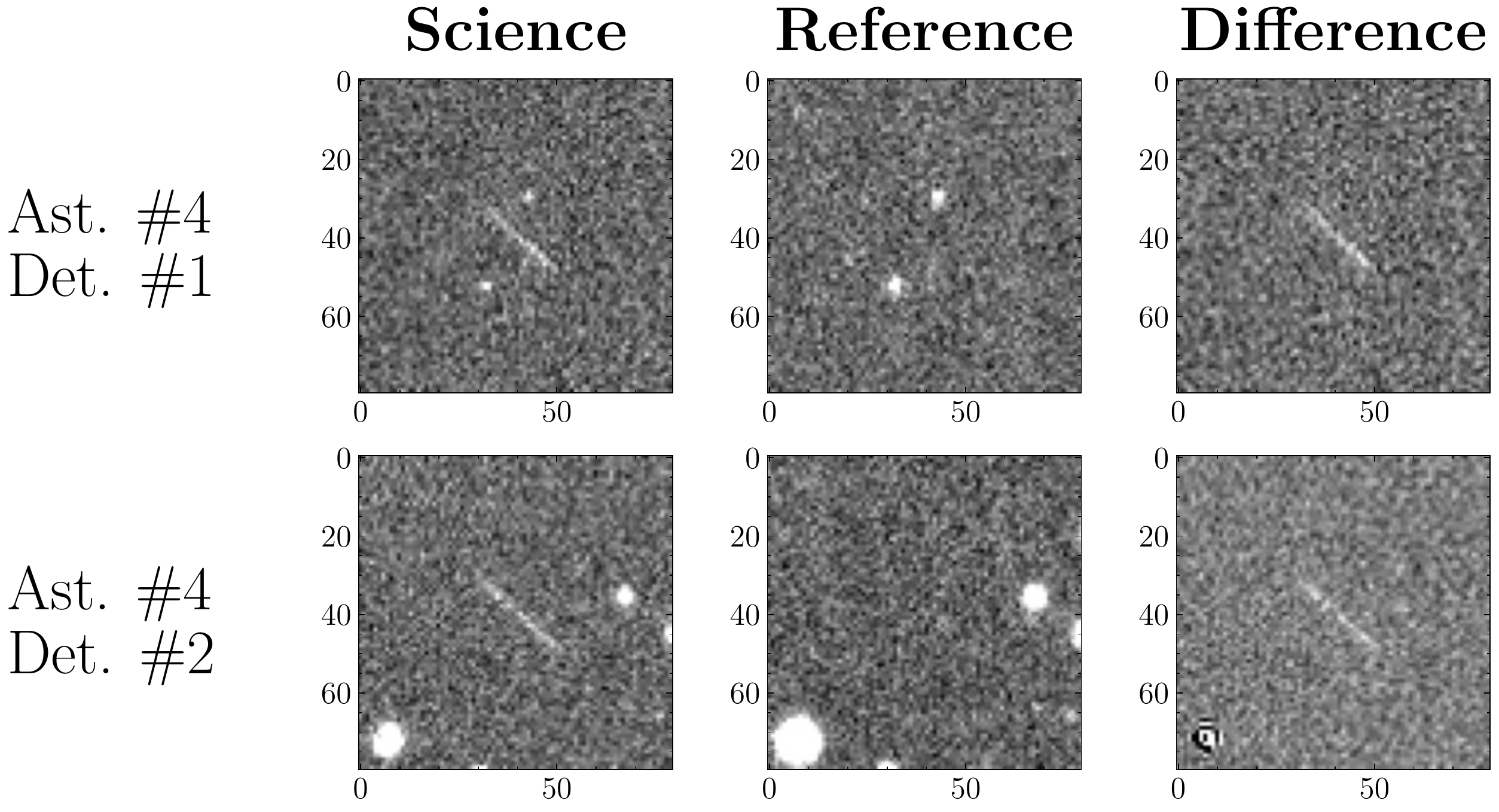}
    \end{minipage}
    \begin{minipage}{0.45\textwidth}
        \centering
        \includegraphics[width=0.95\textwidth]{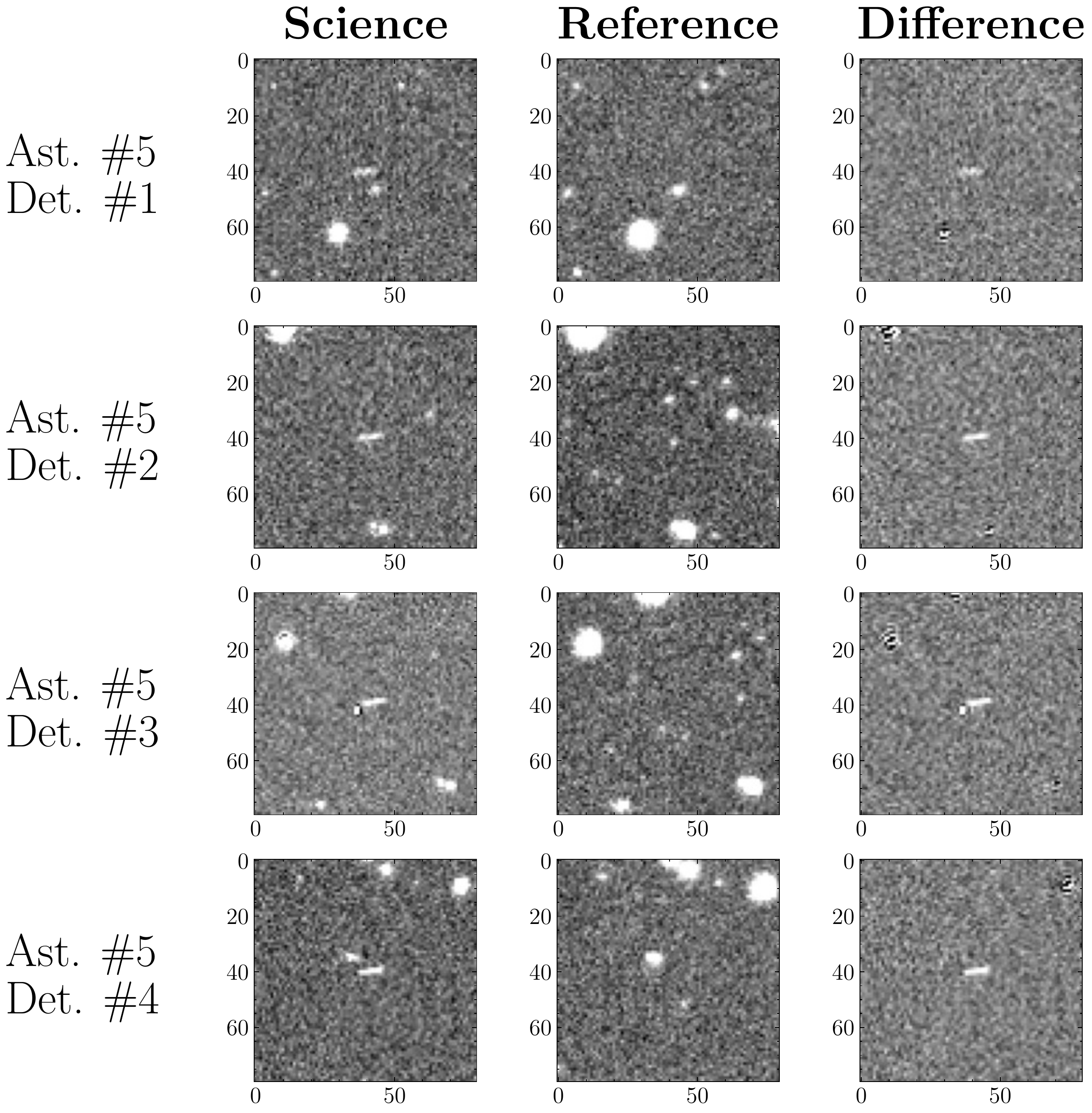}
    \end{minipage}
    \begin{minipage}{0.45\textwidth}
        \centering
        \includegraphics[width=0.95\textwidth]{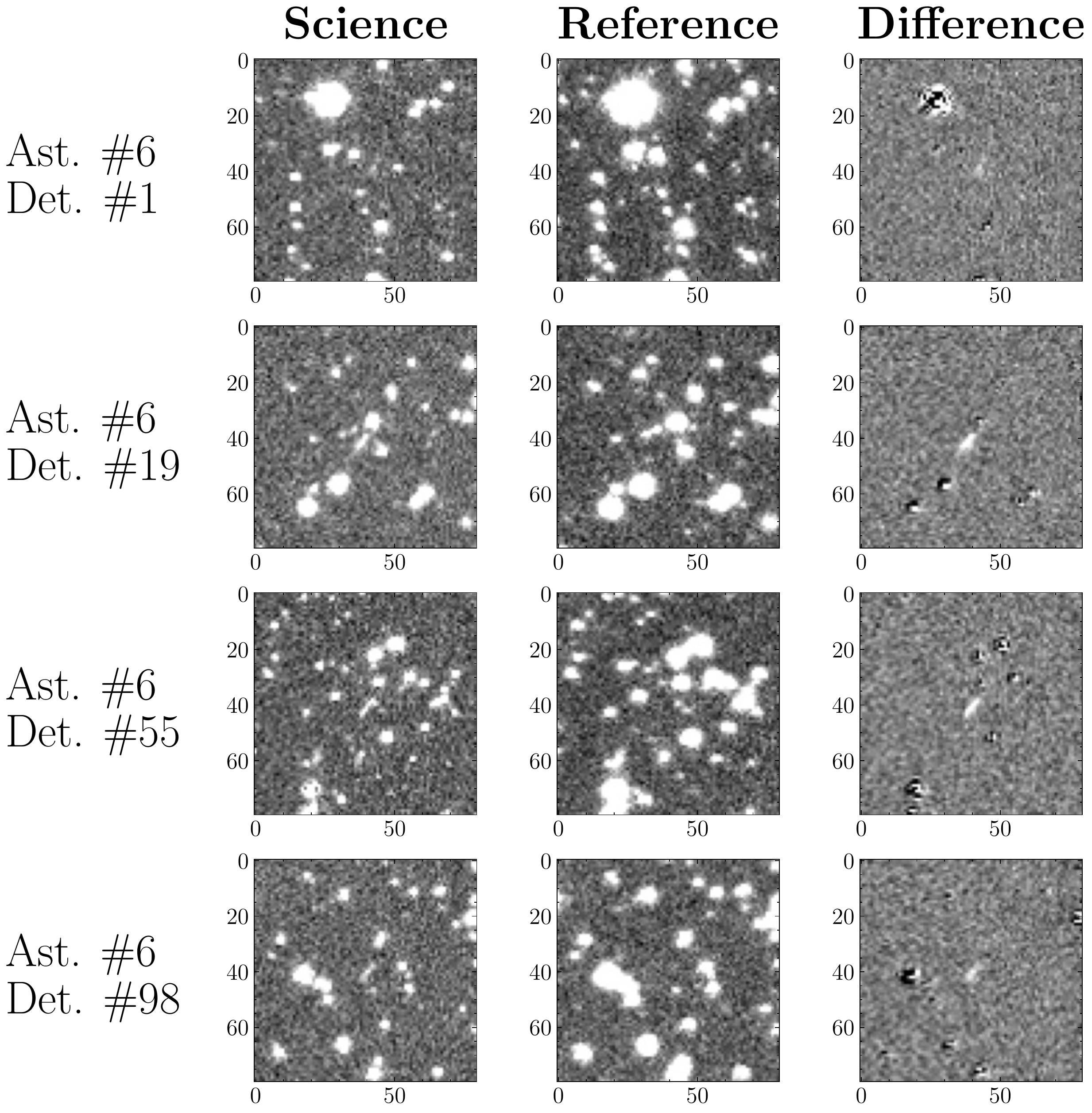}
    \end{minipage}
    
    \caption{Science, reference, and difference image cutouts of our new asteroid discoveries. For asteroid \#2, the difference cutout region for detection \#2 was not available from ZTF. Since we did not have access to the image differencing algorithm that ZTF uses, we were unable to calculate the difference image ourselves. For brevity, we have only included four images of asteroid \#6.}
    \label{fig:asteroids}
\end{figure*}

Our new discoveries of asteroids and detections of previously discovered ones are summarized in Table \ref{tab:summary_dets}. In just four nights of data, we were able to find six new NEOs and detections of five asteroids ZTF and others have discovered previously. Images of these new asteroids are shown in Figure \ref{fig:asteroids}. Note that only two streaked detections of an asteroid are needed to consider it a discovery since each streak contributes two sets of positions and time (one at each endpoint of the streak, which correspond to the start and end of the exposure). This gives us four sets of astrometry which is enough to fit a preliminary orbit.

\begin{table}
\centering
\begin{tabular}{c c c}
\hline
\textbf{Date} & \textbf{Asteroid} & \textbf{\# of Detections} \\ \hline
2019/06/05 & ZTF03XO & 53 \\
\hline
2019/06/06 & New NEO \#1 & 3 \\
 & New NEO \#2 & 2 \\
 & New NEO \#3 & 2 \\
 & ZTF03c6 & 2 \\
 & 2014 MF18 & 3 \\
\hline
2019/06/07 & New NEO \#4 & 2 \\
 & New NEO \#5 & 4 \\
 & 2019 KA4 & 5 \\
 & 2019 LW4 & 6 \\
 & 2014 MF18 & 5 \\
\hline
2019/06/08 & New NEO \#6 & 103
\end{tabular}
\caption{\label{tab:summary_dets} Summary of all detections from each of the four nights.}

\end{table}

\subsection{Streak Discovery Parameters}

\begin{figure*}
    \centering
    \includegraphics[width=0.8\textwidth]{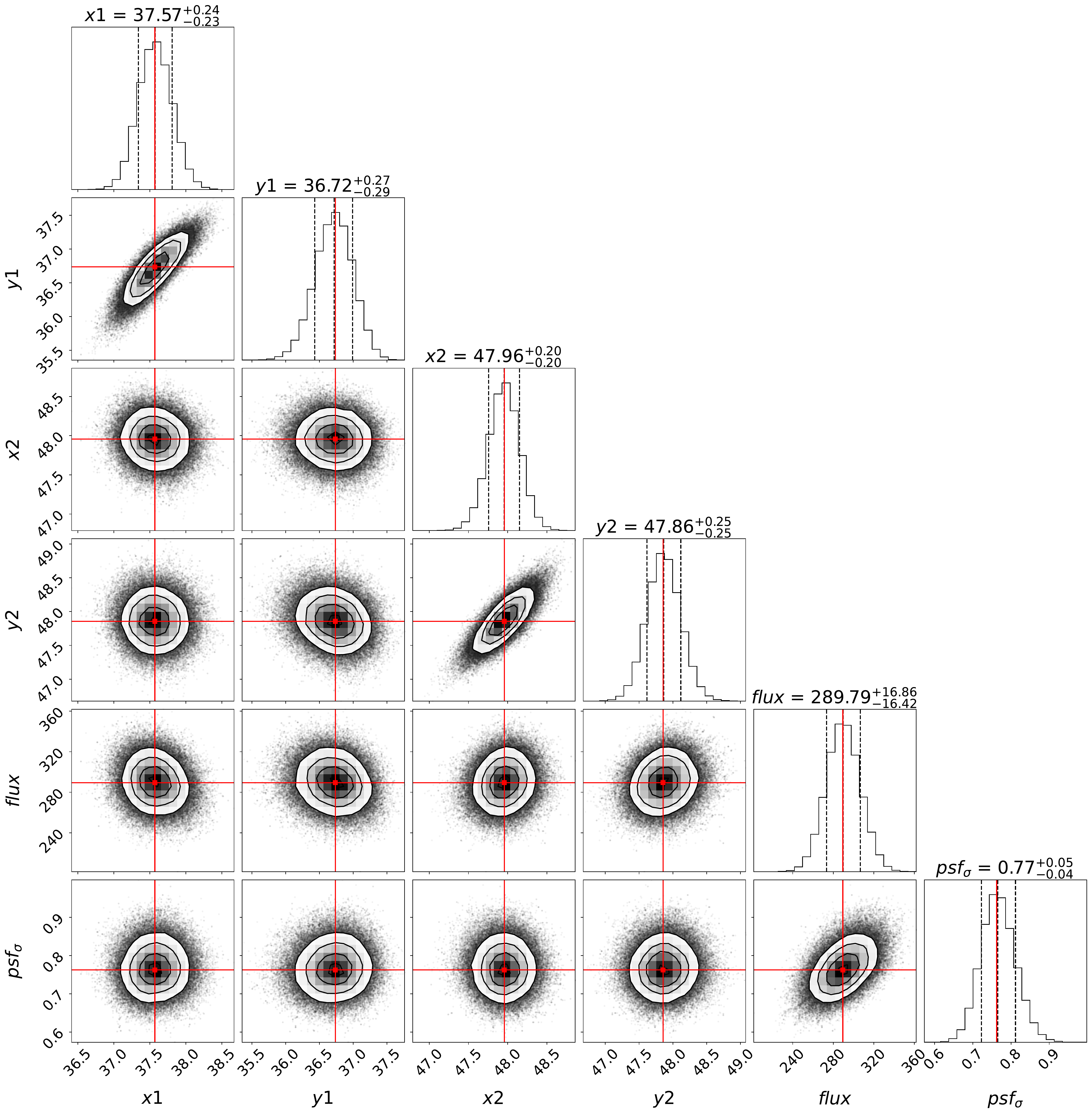}
    \caption{Example of corner plot of the streak fitting parameter distribution for one of our asteroid discoveries (Asteroid \#3, Detection \#1) using the MCMC algorithm. $x1$, $x2$, $y1$, and $y2$ correspond to the coordinates of each endpoint of the streak; \textit{flux} is the net flux of the streak (in ADU); and \textit{psf}$_\sigma$ is the $\sigma$ of the Gaussian PSF used to model the streak. The histograms for each parameter allow us to measure uncertainties using the percentiles. The 16th, 50th, and 84th percentiles are shown as dashed lines while the PDF peak is shown in red.}
    \label{fig:mcmc}
\end{figure*}

For each of our streaks, we fit a Gaussian PSF model (see Equation \ref{eq:flux}) using the Markov chain Monte Carlo (MCMC) method to minimize the mean squared error between the pixels in the difference image and the model. To accomplish this, the approximate locations of the ends of the streak are identified by hand. Then, a pill shaped mask is generated by finding all the pixels within 5 pixels of the line segment connecting the two points. The pixels in the mask are then used to perform the MCMC least-squares fit, which is done using the emcee library \citep{emcee}. Using MCMC allows us to quantify the uncertainties in each of the parameters. An example of a corner plot created from the MCMC sampled streak fitting parameters can be found in Figure \ref{fig:mcmc}.

We also use the mask as a pill aperture for measuring the visual magnitude of our streaks. This is done by directly summing over the difference image (which roughly has a background of 0) pixels that are within the aperture using Equation \ref{equ:vmag}, where $F_{ij}$ refers to difference image pixel intensity (in ADUs), $\text{ZP}$ is the photometric zero point, and $\text{AP\_COR}$ is the visual magnitude correction required for an aperture with a radius of 5 pixels. These photometric parameters are calculated by ZTF and included in headers of the difference image FITS files. The uncertainty calculation is shown in Equation \ref{equ:vmagerr}, where $F_{sci,ij}$ is the science image pixel intensity (which produces the shot noise), $n$ is the number of pixels, $R$ is the read noise (in e$^{-}$), $G$ is the gain of the CCD (in e$^{-}$/ADU), $D$ is the dark current (in e$^{-}$/sec), and $t$ is the exposure time (in seconds).

\begin{align}
    F_{\text{total}}&=\sum F_{ij}\\ \label{equ:vmag}
    V&=-2.5\log_{10}\left(F_{\text{total}}\right) + \text{ZP} + \text{AP\_COR}
\end{align}
\begin{align}
     F_{\text{total,err}}&=\frac{\sqrt{G\sum F_{sci,ij} + nR^2 +nDt}}{G} \\ \label{equ:vmagerr}
    V_{\text{err}}&=\sqrt{\left(\frac{2.5}{\ln{10}}\frac{F_{\text{total,err}}}{F_\text{total}}\right)^2+\text{ZP}_\text{err}^2+\text{AP\_COR}_\text{err}^2}
\end{align}

\begin{figure}
    \centering
    \includegraphics[width=0.4\textwidth]{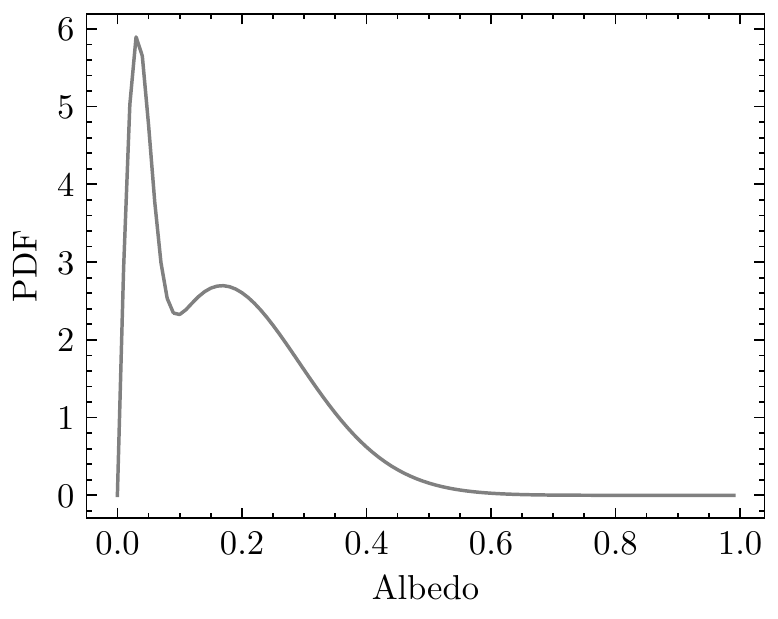}
    \caption{The probability density function for the albedos of near-Earth asteroids from \citet{wrightalbedo}.}
    \label{fig:albedos}
\end{figure}

From the MCMC fits and photometry, we have obtained the astrometry, the length, the $\sigma$ parameter of the Gaussian PSF kernel, and the visual magnitude of each streak. We then use the astrometry and visual magnitude to fit orbits to each of our asteroids. Since our observation arcs are extremely short, there is significant uncertainity in these orbits, and so we use the OpenOrb library, which uses statistical orbital ranging to generate thousands of possible orbits \citep{granvik_openorb}. These orbits can be then sampled to find the probability distributions of the closest approach distances and H magnitudes. The H magnitudes can then be converted to an approximate diameter using the following formula from \citet{bowellastsize}, where $d$ is the diameter, $a$ is the albedo, and $H$ is the H magnitude. This formula assumes that the asteroid is roughly spherical in shape and has a uniform surface. We sample the albedo values from the distrbution derived from WISE data in \citet{wrightalbedo} (see Figure \ref{fig:albedos}).
\begin{equation}
    d=10^{3.1236 - 0.5\log_{10}a - 0.2H}
\end{equation}
\vspace{5pt}

\begin{figure}
    \centering
    \includegraphics[width=0.48\textwidth]{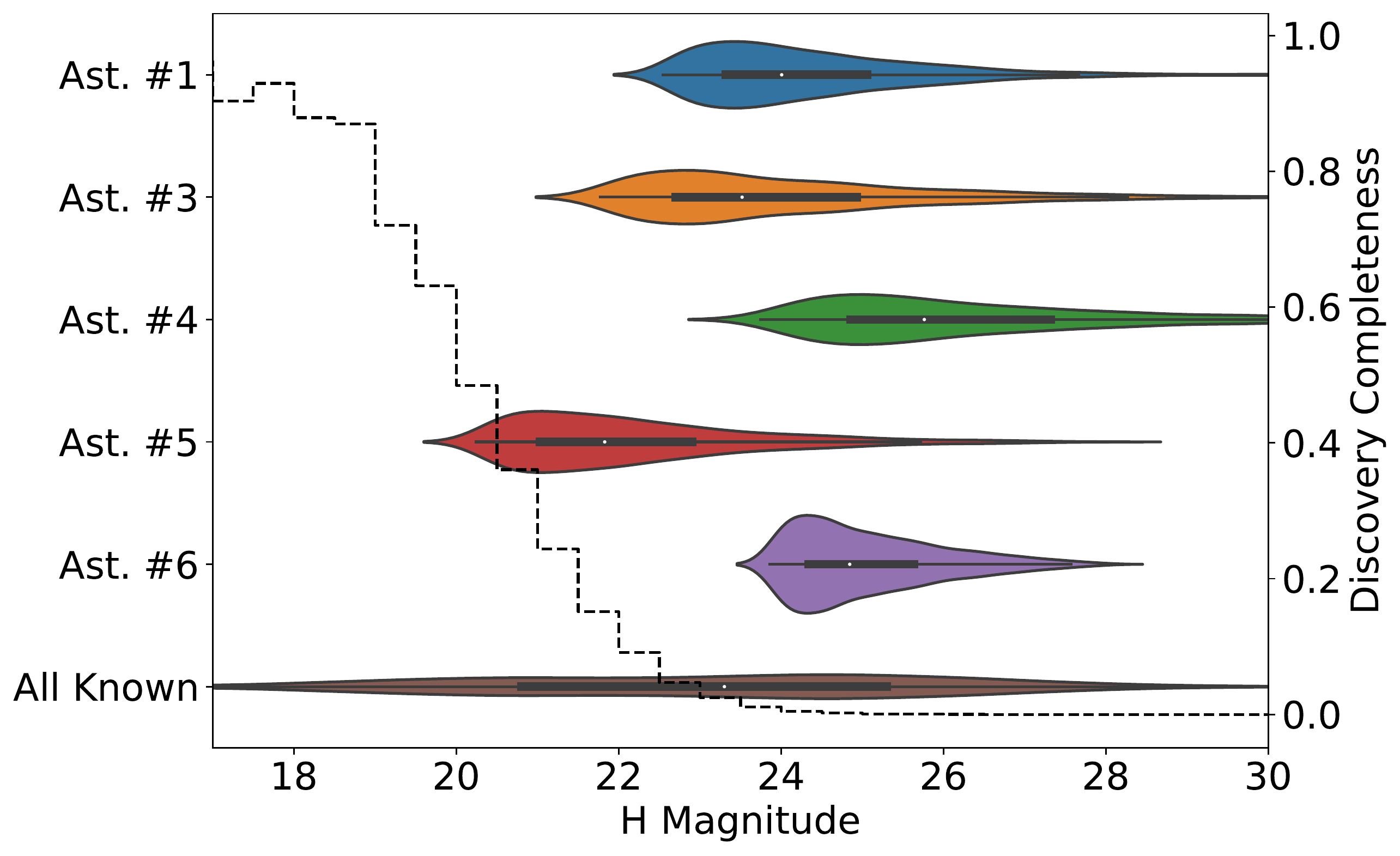}
    \caption{Probability distibution of H-magnitude estimates for each asteroid, where the horizontal axis denotes the H magnitude and the vertical axis denotes the probability density. This is compared against the distribution of the H-magnitudes of all known asteroids (shown in brown) in addition to the discovery completeness for each H-magnitude range (shown as the dotted line). The expected total population of asteroids at a given H-magnitude used to find this discovery completeness were taken from \protect\citet{HARRIS2015302} while the H-magnitudes of known asteroids were taken from the Minor Planet Center (\url{https://www.minorplanetcenter.net/iau/MPCORB/NEA.txt}). Asteroid \#2 is not included due to poor photometric data.}
    \label{fig:hmag_estimates}
\end{figure}

\begin{figure}
    \centering
    \includegraphics[width=0.48\textwidth]{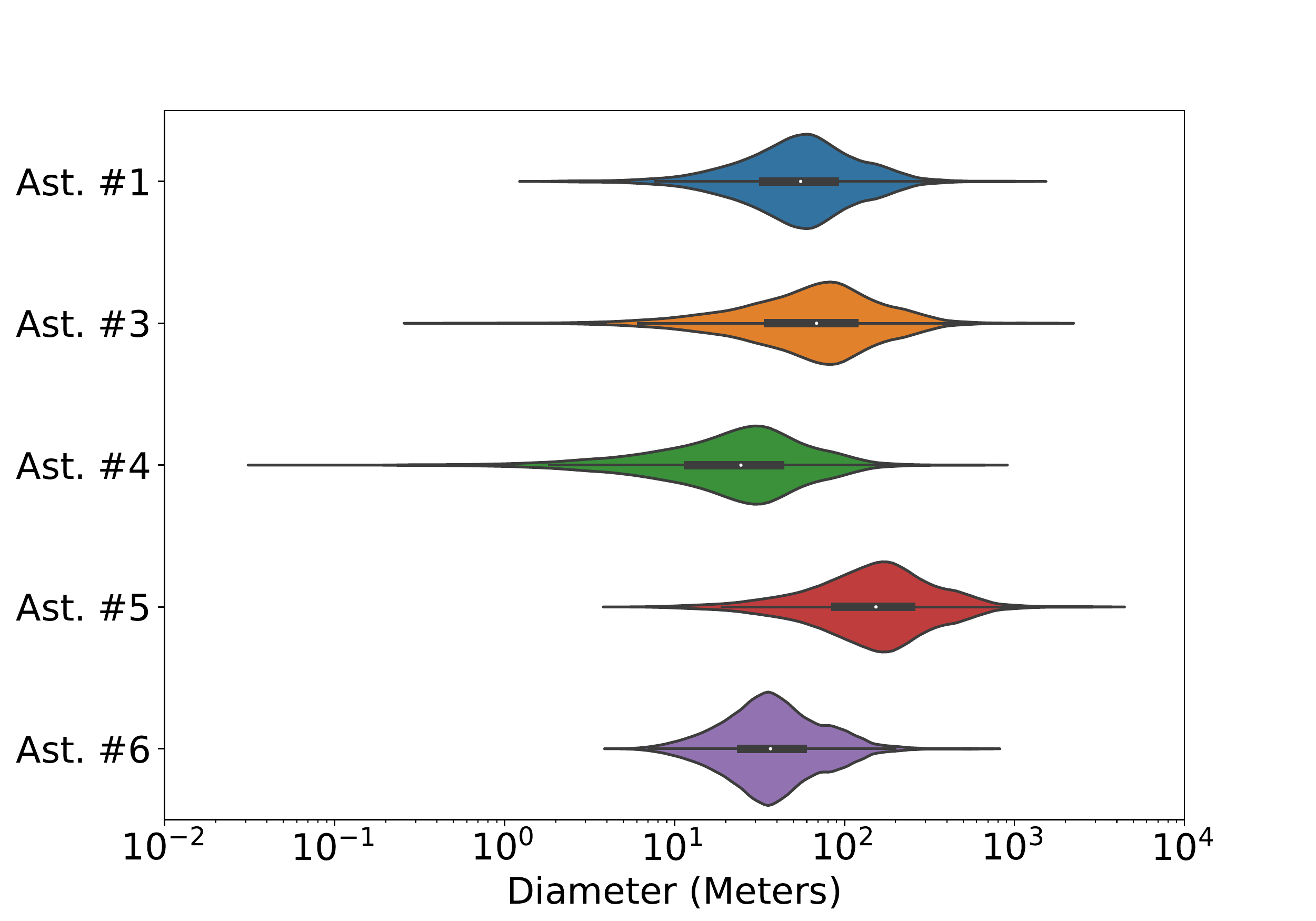}
    \caption{Probability distribution of diameter estimates for each asteroid calculated from the H-magnitude estimates in Figure \ref{fig:hmag_estimates}. The horizontal axis denotes the asteroid diameter (in meters) while the vertical axis denotes the probability density. The albedo is sampled from the distribution found in \citet{wrightalbedo}. Asteroid \#2 is not included due to poor photometric data.}
    \label{fig:diam_estimates}
\end{figure}

\begin{figure}
    \centering
    \includegraphics[width=0.48\textwidth]{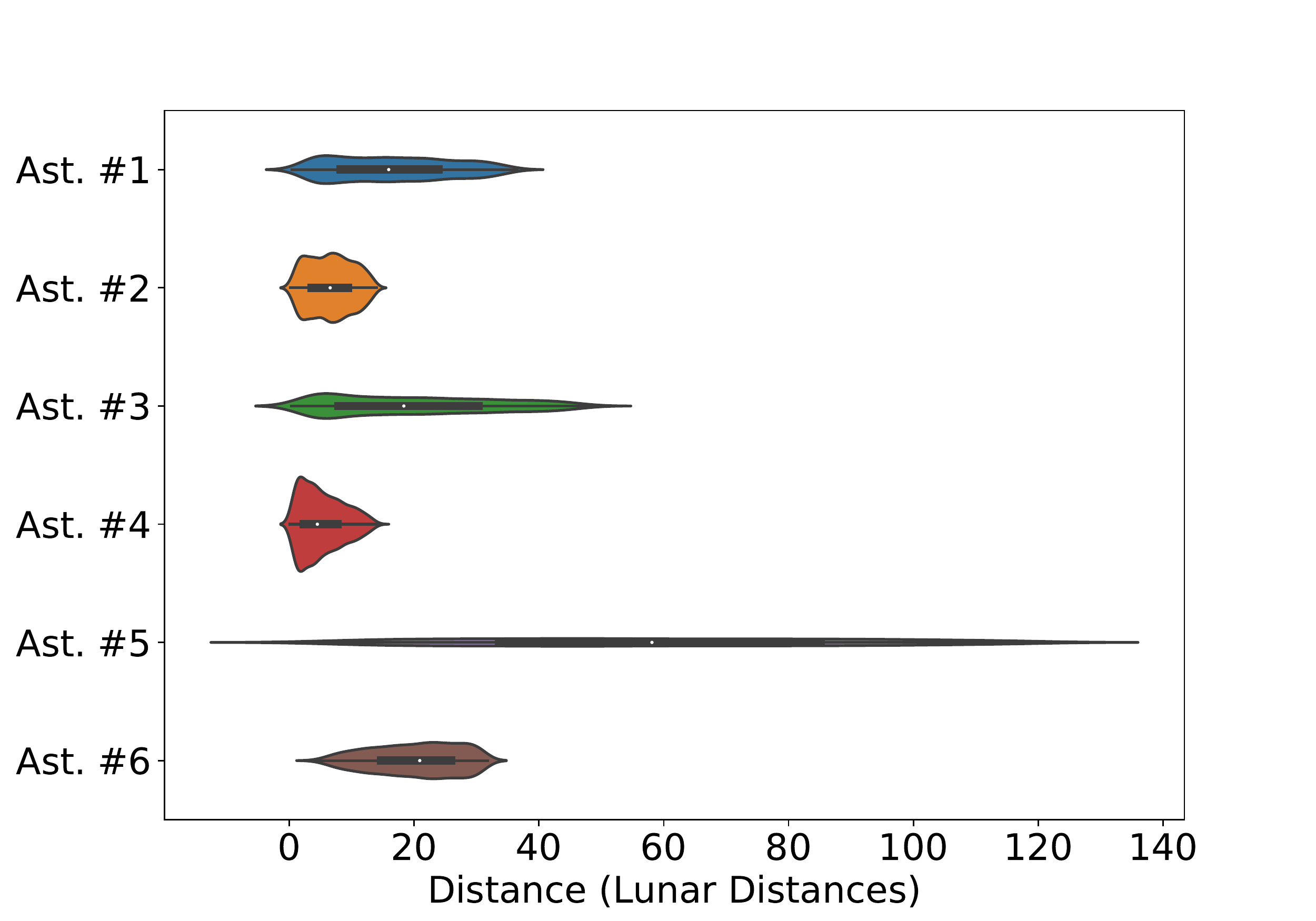}
    \caption{Probability distribution of close approach distances of each asteroid. The horizontal axis denotes the close approach distance (minimum distance to Earth around the time of detection, in lunar distances) while the vertical axis denotes the probability density.}
    \label{fig:dist_estimates}
\end{figure}

The attributes of each new streak detection is shown in Table \ref{tab:params}. The probability distributions of the H-magnitudes, diameters, and close approach distances are also shown in Figures \ref{fig:hmag_estimates}, \ref{fig:diam_estimates}, and \ref{fig:dist_estimates}, respectively. Due to the short observation arcs and the large variability in the possible albedos, the diameter and close approach distances have very high uncertainty. In Figure \ref{fig:hmag_estimates}, we also compare the H-magnitudes of our asteroids to the discovery completeness of asteroids at those H-magnitude ranges. From the figure, we can see that our asteroids mostly correspond to small asteroids with high H-magnitudes that have a low detection completeness. By specifically targeting faint streaks, our algorithm is able to find these smaller asteroids, most of which currently have very low discovery completeness of less than around 10\%.

\renewcommand{\arraystretch}{1.2}
\begin{table*}
\begin{tabular}{l|lllll}
Ast. \#        & Det. \# & Visual Mag & Length             & Diameter (m)             &\multicolumn{1}{c}{Closest Dist. (LD)} \\ \midrule
\multirow{3}{*}{1} & 1            & $19.97\pm0.11$ & $20.5_{-0.9}^{+1.2}$ & \multirow{3}{*}{$57_{-31}^{+59}$}   & \multirow{3}{*}{$16^{+12}_{-10}$}            \\
                   & 2            & $19.32\pm0.05$ & $20.1_{-1.0}^{+1.7}$ &                                   &                                              \\
                   & 3            & $19.25\pm0.06$ & $19.7_{-0.4}^{+0.3}$ &                                   &                                              \\ \midrule
\multirow{2}{*}{2} & 1            & $18.94\pm0.05$ & $20.5_{-0.6}^{+0.7}$ & \multirow{2}{*}{}                 & \multirow{2}{*}{$7^{+4}_{-4}$}               \\
                   & 2            &                &                    &                                   &                                              \\ \midrule
\multirow{2}{*}{3} & 1            & $19.62\pm0.08$ & $15.2_{-0.5}^{+0.5}$ & \multirow{2}{*}{$70_{-45}^{+81}$}   & \multirow{2}{*}{$18^{+17}_{-13}$}            \\
                   & 2            & $18.91\pm0.05$ & $16.6_{-0.4}^{+0.4}$ &                                   &                                              \\ \midrule
\multirow{2}{*}{4} & 1            & $19.15\pm0.05$ & $23.1_{-0.7}^{+0.6}$ & \multirow{2}{*}{$25_{-17}^{+30}$}   & \multirow{2}{*}{$5^{+5}_{-3}$}               \\
                   & 2            & $19.17\pm0.06$ & $29.2_{-2.0}^{+3.4}$ &                                   &                                              \\ \midrule
\multirow{4}{*}{5} & 1            & $20.53\pm0.14$ & $8.2_{-0.4}^{+0.4}$  & \multirow{4}{*}{$157_{-91}^{+168}$} & \multirow{4}{*}{$58^{+38}_{-34}$}            \\
                   & 2            & $19.70\pm0.07$ & $9.0_{-0.3}^{+0.3}$  &                                   &                                              \\
                   & 3            & $19.14\pm0.04$ &                    &                                   &                                              \\
                   & 4            & $19.58\pm0.06$ & $8.5_{-0.2}^{+0.2}$  &                                   &                                              \\ \midrule
\multirow{5}{*}{6} & 1            & $19.40\pm0.06$ & $8.1_{-0.3}^{+0.3}$  & \multirow{5}{*}{$38_{-17}^{+37}$}   & \multirow{5}{*}{$21^{+7}_{-9}$}              \\
                   & 19           & $19.07\pm0.04$ & $7.3_{-0.3}^{+0.3}$  &                                   &                                              \\
                   & 55           & $19.39\pm0.05$ & $7.8_{-0.2}^{+0.2}$  &                                   &                                              \\
                   & 98           & $19.66\pm0.06$ & $7.6_{-0.3}^{+0.3}$  &                                   &                                              \\
                   & 102          & $20.08\pm0.09$ & $7.5_{-0.3}^{+0.3}$  &                                   &                                             
\end{tabular}
\caption{\label{tab:params} Parameters of the detections of our newly discovered asteroid streaks. For the length and size parameters, an upper and lower $\sigma$ is calculated using the 15.9 and 84.1 percentiles to account for the non-normal probability distribution. The length is given in pixels, where each pixel is approximately equal to 1 arcsecond (with an exposure time of 30 seconds). For Asteroid \#2, the difference image from ZTF was masked out for detection \#2, so we were unable to perform a streak fit and calculate a size estimate. For Asteroid \#6, only five detections are shown for brevity.}
\end{table*}

\section{Discussion} \label{sec:discussion}

Using our purely synthetic data-based approach and just four nights of data, we were able to find 6 new NEOs which other observatories, including ZTF, did not find, in addition to several NEOs that other observatories have already discovered. This is the first ever successful use of a machine learning algorithm that does not necessitate the use of manual collection of training data in order to detect new asteroid streaks, thus overcoming the limitations that manually annotated data has. By having to use manually collected data to train a CNN, observatories are limited to a training dataset proportional to the size of their collected data, they can not specifically target dim streaks or streaks of a certain length, and they have to spend lots of time hand labeling each image. This research shows that synthetically generated streaks can accurately match the appearance of real asteroid streaks to the point where a CNN can be only trained on synthetic streaks and still be able to detect real undiscovered asteroids. 

Applying our machine learning model to past streaks detected by ZTF, we are able to attain extremely high detection completeness (97.4\% for confirmed and 95.6\% for unconfirmed asteroids) which nearly matches the 97\% true positive rate of our simulated data (Figure \ref{fig:detcompltdets}). We theorize that the lower detection completeness for the unconfirmed asteroids is possibly due to the shortage of long streaks in our dataset as the majority of streaks we generate are less than 20 pixels in length. As shown by the missed detections in Figures \ref{fig:failedconf} and \ref{fig:failedunconf}, it seems that some longer streaks have the potential to be missed by our CNN. Another major issue was streaks that experience heavy occlusion from other stars, such as the last streak in Figure \ref{fig:failedunconf}. The difficulty in detecting these occluded streaks likely lies in how we remove simulated streaks which overlap a star significantly to reduce diffraction spike detections. If we further improved our algorithm to find occluded streaks, it is possible that we would detect more diffraction spike false positives. In the future it may be beneficial to try to simulate diffraction spike negative samples rather than remove occluded simulated streaks.

\begin{figure}
    \centering
    \includegraphics[width=0.48\textwidth]{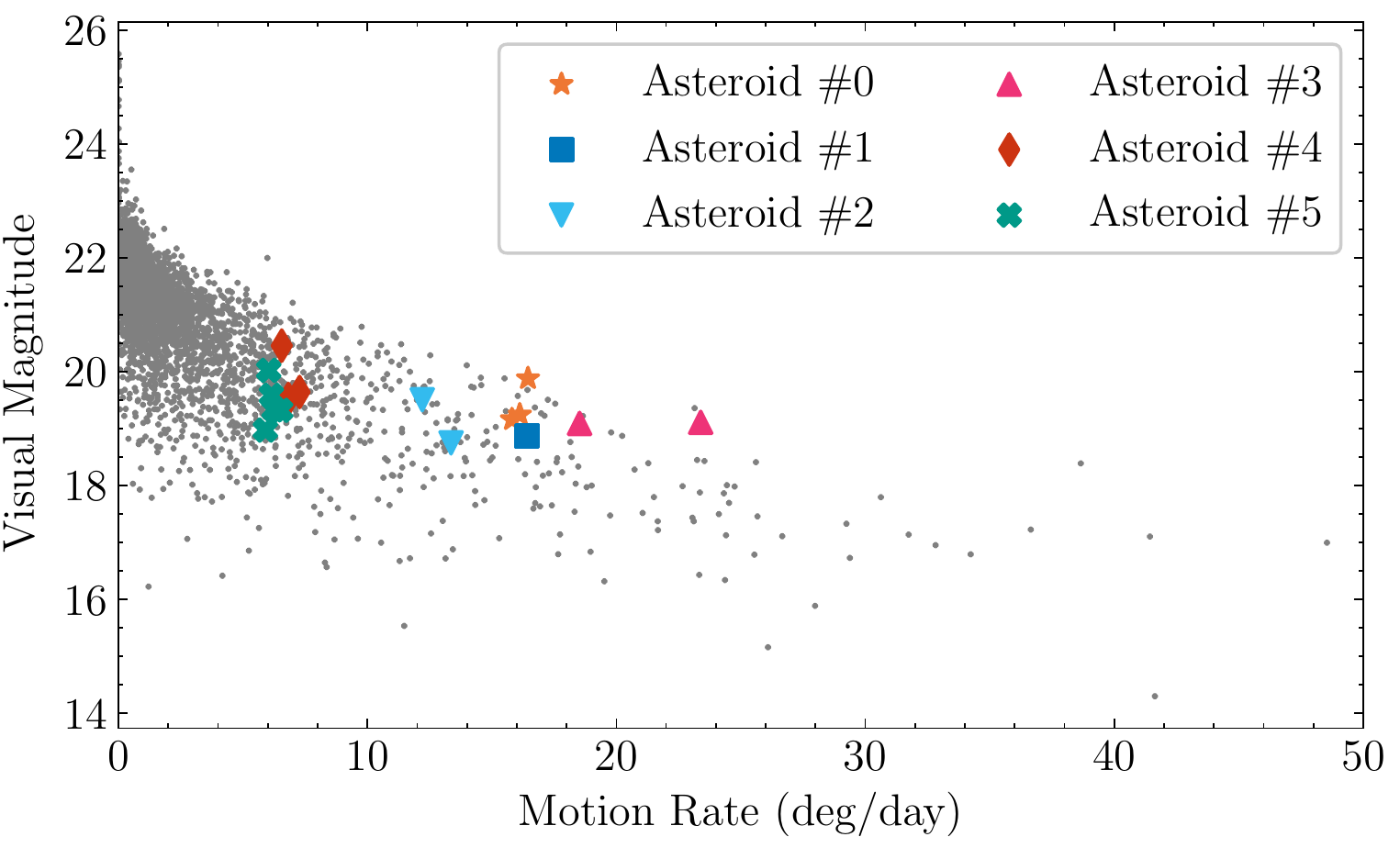}
    \caption{Comparison of asteroid discovery circumstances in the 2020 MPECs to our asteroids.}
    \label{fig:comp_prev_discov}
\end{figure}

\begin{figure}
    \centering
    \includegraphics[width=0.48\textwidth]{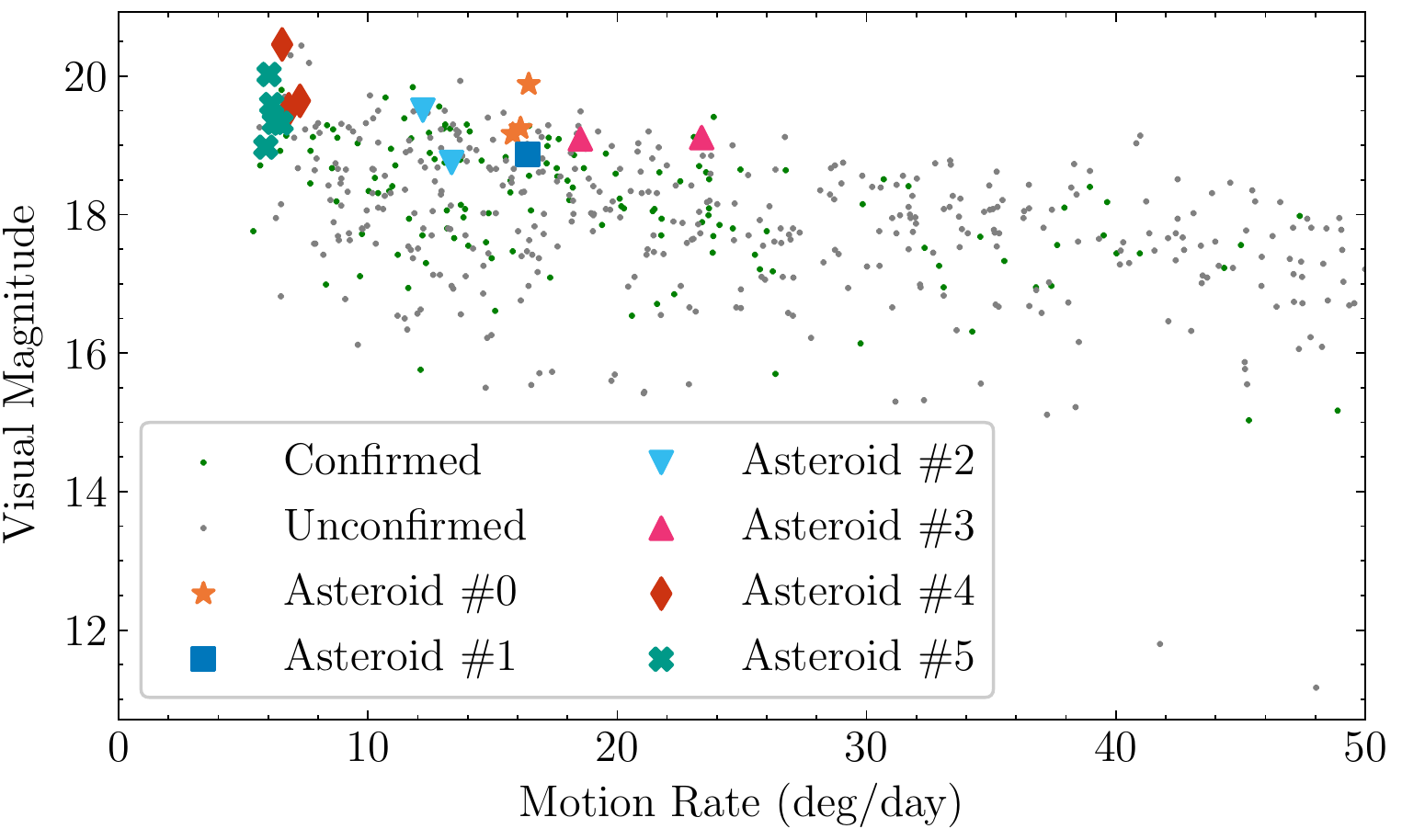}
    \caption{Comparison of all confirmed and unconfirmed (observation arc too short for MPC to assign a designation) ZTF asteroid streak discoveries to our discoveries.}
    \label{fig:comp_prev_ztf}
\end{figure}

We collected motion rate and visual magnitude data from all the Minor Planet Electronic Circulars (MPECs) of 2020, which consist of discoveries of unusual minor planets -- primarily near-Earth Objects. This data is plotted against our discoveries in Figure \ref{fig:comp_prev_discov}. We only use the initial discovery detections from the MPECs so that we can specifically compare discoveries as opposed to follow-up observations of known objects. We also created a similar graph comparing our discoveries to ZTF's, both confirmed and unconfirmed by the MPC (Figure \ref{fig:comp_prev_ztf}). From both these figures, we can see that relative to other asteroids travelling at similar speeds, our asteroids tend to be somewhat fainter. By focusing the brightness distribution of streaks in our dataset on fainter asteroids, the machine learning model is able to robustly detect these faint objects. Compared to the MPEC results, our asteroids tend to be on the faster side as most NEOs are detected as slow moving, point-source objects. However, compared to ZTF discoveries, our asteroid streaks are on the shorter side but still fainter, suggesting that our model may be effective at detecting short and faint asteroid streaks.


Although we largely used parameter distributions of real asteroids to create artificial training datasets, these datasets are likely biased towards those asteroids which could be easily detected by previous methods. Our training samples and finally detected samples are likely biased towards these asteroids. In order to have a more complete detection of streaks (such as long streaks) with our method, training samples with large weights towards rare groups are necessary to help detect them. 

\subsection{Limitations and Drawbacks}
Because archival data was used in our search for NEOs, we were unable to obtain the follow-up observations required for most of our streak detections to be confirmed by the Minor Planet Center (MPC). If we had access to more recent data, it is possible that other observatories would have done follow-up observations of these streaks, satisfying the MPC's 24 hour observation arc requirement. However, we were able to find additional observations of Asteroid $\#$6 on 2019/06/09, partly due to the asteroid being located in a high cadence field which allowed us to obtain over 100 observations on the first night (see \citet{kupferztfcadence} for more details). This NEO has now been given the designation 2019 LH27 by the MPC. In the future we plan to continue exploring high cadence data from ZTF's archives to find asteroids which may potentially appear across multiple nights in ZTF data.

In this work, we only focused on shorter and fainter streaks. However, if we expanded our simulated dataset to include more bright and long streaks we likely would have better detection completeness for the ZTF discoveries we missed, as described in Section \ref{sec:cnnperformance}. For longer streaks however, it may become necessary to model the brightness fluctuations of asteroids due to rotation as small asteroids tend to have very short rotation periods on the order of minutes that would become more pronounced with longer streaks \citep{hergenrother2011}.

While using the science and reference images rather than the difference images potentially allows our model to have access to more information, our normalization algorithm can cause bright pixels to be clipped, making it difficult to detect streaks significantly overlapping with bright stars. Because of this, it may be helpful to provide science, reference, and difference images to the machine learning model, as the difference images would subtract out stars that are overlapping with a streak. Increasing the clipping range to be larger than -5 to 5 could also allow for better detections. 

It is important to note that ZTF's algorithm uses a different preprocessing algorithm for extracting potential streak candidates to be fed into their CNN model. Their algorithm focuses on longer asteroid streaks and also has stricter thresholds to reduce false positives. Because of this difference in preprocessing, our CNN and synthetic training approach is not necessarily \textit{better} at streak detection compared to ZTF's algorithm. In particular, the 5th and 6th asteroids which we found were likely missed by ZTF since they are shorter than the streaks they tended to focus on during that time period of the survey. Nevertheless, the fact that we are able to recover most of ZTF's discoveries using no real image data for the training set and are also able to detect undiscovered objects, suggests that our approach allows for the very robust detection of asteroid streaks based on the distributions that we target our dataset towards. Our work has shown that real image data is not necessary to train an effective asteroid streak detection system and that a purely simulated dataset may allow for the detection of asteroid streak distributions like faint streaks or short streaks for which detection completeness is low.

Given different strengths for both algorithms, we would suggest to ZTF that our pipeline could be a potential supplement which runs in addition to their current DeepStreaks algorithm by targeting specific distributions of streaks which may be harder to detect. Moreover, we believe the main benefit of our algorithm is derived from the removal of the need for real asteroid streak data collection, which allows other surveys besides ZTF to much more easily adopt streak detection.

\subsection{Future Potential Applications}

As all the data we use to train the model is simulated, this approach can be more easily applied to other transient surveys than approaches that require real data. For instance, ZTF is a prototype for the future Vera C. Rubin Observatory. Applying our method to the Rubin Observatory, which has a higher limiting magnitude of roughly 24.7 for the red band filter as opposed to 20.4 for ZTF, fainter asteroids would appear at a higher SNR, making it much easier for our algorithm to detect them. Quantitatively, we can compare the \'etendue and volumetric survey speed\footnote{The volumetric survey speed roughly quantifies the volume of space where an object of a certain absolute magnitude can be detected divided by the time per exposure, allowing us to take into account survey speed (see \citet{Bellm_2016} for details)} of both telescopes. The Rubin Observatory has an \'etendue of $319.5\text{ m}^2\text{ deg}^2$ and a volumetric survey speed of $3.7 \times 10^5 \text{ Mpc}^3\text{/s}$ while ZTF only has an \'etendue of $53.1\text{ m}^2\text{ deg}^2$ and a volumetric survey speed of $2.5 \times 10^4 \text{ Mpc}^3\text{/s}$ \citep{Bellm_2016}. Using our method, faint asteroid streaks can be more robustly detected, allowing us to take advantage of the increased detection capabilities to the fullest extent possible.

\section*{Acknowledgements}

Based on observations obtained with the Samuel Oschin 48-inch Telescope at the Palomar Observatory as part of the Zwicky Transient Facility project. ZTF is supported by the National Science Foundation under Grant No. AST-1440341 and a collaboration including Caltech, IPAC, the Weizmann Institute for Science, the Oskar Klein Center at Stockholm University, the University of Maryland, the University of Washington, Deutsches Elektronen-Synchrotron and Humboldt University, Los Alamos National Laboratories, the TANGO Consortium of Taiwan, the University of Wisconsin at Milwaukee, and Lawrence Berkeley National Laboratories. Operations are conducted by COO, IPAC, and UW.

This research has made use of data and/or services provided by the International Astronomical Union's Minor Planet Center.

The authors would like to thank Quanzhi Ye and Bryce Bolin for their helpful discussions and the anonymous referee who has provided valuable suggestions which helped improve the quality of this paper.

\section*{Data Availability Statement}
The data underlying this article were accessed from the Zwicky Transient Facility's public data releases (\url{https://www.ztf.caltech.edu/}). The derived data generated in this research will be shared on reasonable request to the corresponding author.

\bibliographystyle{mnras}
\bibliography{mnras_ref} 

\bsp
\label{lastpage}
\end{document}